\documentclass[a4paper,fleqn,usenatbib]{mnras}

\usepackage{mathptmx}
\usepackage[T1]{fontenc}
\usepackage{ae,aecompl}

\usepackage{graphicx}	% Including figure files
\usepackage{caption}
\usepackage{amsmath}	% Advanced maths commands
\usepackage{amssymb}	% Extra maths symbols

\title[Mixing and overshooting in DA white dwarfs]{Mixing and Overshooting in Surface Convection Zones of
DA White Dwarfs: First Results from ANTARES
}

\author[F. Kupka et al.]{
F. Kupka,$^{1,2,3}$\thanks{E-mail: friedrich.kupka@univie.ac.at}
F. Zaussinger,$^{4}$
and M.H. Montgomery$^{5,6}$
\\
$^{1}$Wolfgang Pauli Institute, c/o Faculty of Mathematics, Univ.\ of Vienna, Oskar-Morgenstern-Platz 1, A-1090 Wien, Austria\\
$^{2}$Faculty of Mathematics, Univ.\ of Vienna, Oskar-Morgenstern-Platz 1, A-1090 Wien, Austria\\
$^{3}$Institute for Astrophysics, Faculty of Physics, Univ. G\"ottingen, Friedrich-Hund-Platz 1, D-37077 G\"ottingen, Germany\\
$^{4}$Dept. of Aerodynamics and Fluid Mechanics, Brandenburg University of Technology Cottbus-Senftenberg, Germany \\
$^{5}$Department of Astronomy and McDonald Observatory, University of
Texas at Austin, Austin, TX, 78712, USA\\
$^{6}$Center for Astrophysical Plasma Properties, University of Texas
at Austin, Austin, TX, 78712, USA
}

\date{Accepted XXX. Received YYY; in original form ZZZ}

\pubyear{2017}

\begin{document}
\label{firstpage}
\pagerange{\pageref{firstpage}--\pageref{lastpage}}
\maketitle

\begin{abstract}

  We present results of a large, high resolution 3D hydrodynamical
  simulation of the surface layers of a DA white dwarf (WD) with
  $T_{\rm eff}=11800$~K and $\log(g)=8$ using the ANTARES code, the
  widest and deepest such simulation to date. Our simulations are in
  good agreement with previous calculations in the
  Schwarzschild-unstable region and in the overshooting region
  immediately beneath it. Farther below, in the wave-dominated region,
  we find that the rms horizontal velocities decay with depth more
  rapidly than the vertical ones. Since mixing requires both vertical
  and horizontal displacements, this could have consequences for the
  size of the region that is well mixed by convection, if this trend
  is found to hold for deeper layers. We discuss how the size of the
  mixed region affects the calculated settling times and inferred
  steady-state accretion rates for WDs with metals observed in their
  atmospheres.
  
 \end{abstract}


%%%%%%%%%%%%%%%%%%%%%%%%%%%%%%%%%%%%%%%%%%%%%%%%%%

%%%%%%%%%%%%%%%%% BODY OF PAPER %%%%%%%%%%%%%%%%%%

%
\section{Introduction}

\citet{siedentopf33r} first suggested that the surface layers of white
dwarfs should be convective. 
% A more precise spectroscopic distinction
% of different types of white dwarfs and more accurate determinations of
% their surface gravities were yet to come, but this 
This insight still holds today: white dwarfs of type DA, 
characterised by a surface composed of (nearly) pure hydrogen,
have convection zones due to the partial ionization of hydrogen for
effective temperatures $T_{\rm eff} \lesssim 14000$~K at a surface gravity of $\log(g)=8$. 
% For higher or lower values of $\log(g)$ this upper boundary shifts to higher and
% lower values of $T_{\rm eff}$, respectively. As in sufficiently cool
% main sequence stars and giants, these convection zones range deeper
% from the lower photosphere into the stellar envelope in case of a
% lower effective temperature and their depth has a similar dependence
% on surface gravity which is also observed for the just mentioned
% ``convective boundary''. 
Theoretical calculations confirming this
qualitative picture include envelope and evolutionary models of white
dwarfs \citep[e.g.,][]{vanHorn70,fontaine76} as well as numerical,
hydrodynamical simulations of limited regions of the stellar surface
(``box-in-a-star'' calculations) in 2D \citep[two spatial
dimensions,][]{freytag96b} and more recently in 3D \citep[three
spatial dimensions,][]{tremblay11b,tremblay13b,tremblay15b}. Such 2D and 3D
simulations also allow the determination of mixing below the
convective zone, similar to non-local Reynolds stress models
\citep{montgomery04b}.

In this work, we address the question of overshooting and mixing
induced by this {\em surface convection zone} of DA white dwarfs (DA
WDs) in the context of accretion and diffusion processes.  A sizable
fraction ($\sim\,25\,$\%) of DA WDs show evidence of metal lines in
their spectra \citep{gianninas14}. Since the theoretical gravitational
settling times \citep[from days to thousands of
years,][]{koester06,koester09} are much shorter than the evolutionary
times, this is taken to be evidence of ongoing or recent accretion.
In order for charge neutrality to be maintained by a plasma in a
gravitational field, a weak electric field is set up; this electric
field leads to a settling velocity of the trace amounts of metals in a
predominantly hydrogen background \citep{arcoragi80}. Since the
velocities in the convection zone are much larger than the computed
settling velocities, the convective region acts as a single zone, and
the settling time for the surface abundances is determined by the
settling time at the point beneath the convection zone at which the
velocities from the convection zone produce negligible mixing compared
to gravitational settling \citep[see][]{dupuis92}.

With the advent of detailed 3D numerical simulations of stellar
convection zones, it becomes possible to compute, in principle, the
amount of mass mixed in the ``thin'' surface convection zones of DA
WDs. These convection zones are just a few km deep and contain a very
small fraction of the total stellar mass; for DA WDs with $\log(g)=8$
and $T_{\rm eff} \sim 11500~{\rm K}$, \citet{tremblay15b} derive the
mass of the convection zone to be $\sim 10^{-14}$ of the total stellar
mass.

We analyse a 3D numerical simulation calculated with ANTARES
\citep{muthsam10b} that differs from previous work by considering a
computational box much wider and much deeper such that a larger
fraction of the stellar envelope mixed by convective overshooting is
contained inside it. We discuss models for the extent of overshooting
underneath convection zones in this class of objects, in particular
the so-called {\em exponential overshooting} suggested first in
\citet{freytag96b} in the context of A-type stars and proposed by
\citet{herwig00b} to be applied to a much larger variety of objects
(for white dwarfs cf.\ \citealt{tremblay15b}).  The simulation data
are then used to infer the extent of the convectively mixed region and
to investigate the effect this mixing has on the inferred settling and
accretion rates of metals in DA WDs.

In Sect.~\ref{sec:numsim} we describe model parameters and procedures
of relaxation and statistical evaluation for our numerical
simulation. Its mean thermal structure and the velocity fields are
evaluated in Sect.~\ref{sec:numres}.  We analyse these data with
respect to convective mixing and calculate the mixed mass in
Sect.~\ref{sec:analysis}, followed by a discussion and an outlook in
Sect.~\ref{sec:discussion}.

\section{Numerical simulation}  \label{sec:numsim}
Since surface convection zones of DA WDs are very shallow compared to the 
stellar radius, we can use a box-in-a-star ansatz and confine our numerical 
simulation to a small volume located at the stellar surface. As the stellar 
photosphere is optically thin we need to solve the radiative transfer equation to 
compute the radiative flux in the upper part of the simulation volume. Since 
the DA WDs are strongly stratified with high enough velocities to produce shock 
fronts (cf.\ the root mean square velocities published by \citealt{tremblay13b,tremblay15b}),
the fully compressible conservation laws of radiation hydrodynamics
have to be solved numerically. This calculation is performed with the ANTARES software suite
which has been utilised for various astro- and geophysical applications. The main
purpose of the code is the simulation of the solar surface convection zone \citep{muthsam10b,grimm-strele15c}. 
However, several add-ons extended the code for convection simulations of other types of stars, as e.g. A-stars \citep{Kupka_2009} or  
Cepheids \citep{Mundprecht_2013}. Stellar interior simulations concerning semi-convection have been investigated 
by \citet{Zaussinger_2013a} for the fully compressible and incompressible formulation, and subsequently for the general
Mach number regime by \citet{Happenhofer_2013}.

We first describe our setup for the simulation code ANTARES 
\citep{muthsam10b} which we use to solve the governing radiation hydrodynamic 
equations on a three-dimensional rectangular Cartesian grid ($x$-coordinate vertical, 
$y$ and $z$ ones horizontal). The radiative heating and cooling of gas is modelled by 
computing the radiative heat exchange rate $\rm Q_{\rm rad}$ from solving the radiative 
transfer equation by a short characteristics method in the grey approximation. Below an optical 
depth of $\rm \tau\approx 150$, i.e., for the lower $\rm 75\%$ of the simulated region, 
the diffusion approximation is used instead. The equation of state for the pure hydrogen 
WD is given by a tabulation from the OPAL database \citep{rogers96b} for $Z=0$. Rosseland opacities 
$\kappa_{\rm ross}$ for pairs of $\rm (\rho,T)$ points (density and
temperature, respectively) are given by \citet{iglesias96b}. 
Fluid can leave and enter through the top vertical layer located in the upper photosphere \citep{grimm-strele15b}. 
The lower vertical boundary is located deeply inside the radiative region and is thus assumed 
to be closed with vertically stress-free conditions for the horizontal 
velocities and a radiative flux $F_{\rm rad}=F_{\rm total}= F_{\rm input} \equiv \sigma T^4_{\rm eff}$ 
entering the box such that $T_{\rm eff}=11800$~K. Periodic boundary conditions are
assumed along horizontal directions. We applied only small initial density fluctuations directly 
inside the convectively unstable zone to trigger convection to minimise relaxation time particularly
of the lower radiative zone. The high mixing rate of the downdrafts rapidly damps initial patterns 
and guarantees statistically unbiased data. The ANTARES code is parallelised in a hybrid way 
in MPI and OpenMPI, and we have used up to 576 cores for this simulation.

\begin{figure}
	\includegraphics[width=\columnwidth]{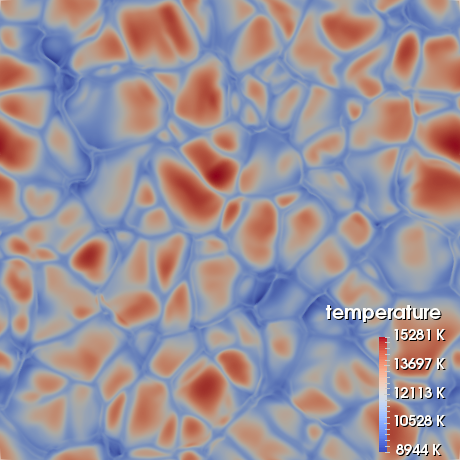}
    \caption{Temperature field for a horizontal cut where the horizontally averaged temperature 
       $\langle T \rangle \approx {\rm T_{\rm eff}}$, at 912~m below the top of the simulation box,
       for a snapshot of our 3D simulation. 
       % Root mean square fluctuations of $T$ are less than 10\% for both the snapshot
       % (1170 K) and its average over $t_{\rm stat}$ (1128 K).
       }
    \label{fig:granules}
\end{figure}

Our starting model has $T_{\rm eff}=11800$~K, a gravitational 
acceleration of $\rm \log(g)=8$, and a pure hydrogen composition.
Vertical extent and numerical resolution are determined by the effective 
height of the convective zone {\em including} the overshooting 
region and by requiring a well-resolved thermal structure. The horizontal 
directions scale with an aspect ratio of $\rm 2.2$ and allow for 
$\gtrsim 6\text{--}8$ granules in each direction. A vertical resolution 
of $\sim 15$ grid cells per pressure scale height $H_{\rm p}$ ensures 
an average change of pressure $\lesssim 7\%$. For a total vertical extent 
of $\sim 16.8\, H_{\rm p}$, $\rm 252$ cells are used vertically while 
$\rm 522 \times 522$ cells are used horizontally, with grid spacings 
$\rm \Delta x=29.42\,m$ vertically and $\rm \Delta y = \Delta
z=31.19\,m$ horizontally. The
simulated box thus has a volume of $\rm 7.384\times 16.28\times 16.28\, {\rm km}^3$
and consists of about $6.8\cdot10^7$ grid cells.  
A snapshot from this simulation is shown in Fig.~\ref{fig:granules}.

We use the time a sound wave requires to travel from top to bottom as
a unit here, whence $\rm \tau_{scrt}=0.236\,sec$. The simulation 
has been run for $\rm 92\,scrt$,
i.e., $\rm 21.7\,sec$, resulting in $\rm 2\,TB$ of data. 
This includes $t_{\rm 2D}=\rm 30\,scrt$ of initial relaxation of the mean 
structure by a 2D simulation with otherwise identical extent, resolution, 
and physical parameters. The 2D simulation was started from 
a 1D model computed with the Warsaw envelope code
\citep{paczynski69,paczynski70,pamyatnykh99} and saves relaxation time
for the mean stratification in the overshooting zone which cannot be
accurately guessed from our 1D model. The transfer from 2D to 3D was
established by horizontally averaging a snapshot of the 2D simulation
to generate a new, `pre-relaxed 1D model' which was then used as
initial condition for the 3D simulation. The time $t_{\rm 2D}$ was
chosen to ensure zero total vertical momentum at a time where strong
vortices in the overshooting zone have not yet developed. The
pre-relaxed 1D model was then perturbed again as described above.
The statistical analysis is based on snapshots of density, momentum, 
internal energy, radiative flux, and pressure made each $\rm 0.1\,scrt$. 
Further quantities of interest can be calculated in post-processing,
e.g., mean values (horizontally and also in time), variance, 
skewness, kurtosis, and cross correlations of various fields. 
The statistical analysis has been performed from averages over 
the last $t_{\rm stat} \sim \rm 22\,scrt$ of the simulation.

\section{Mean structure, relaxation, and velocities}    \label{sec:numres}

\subsection{Mean structure}

\begin{figure}
	\includegraphics[width=\columnwidth]{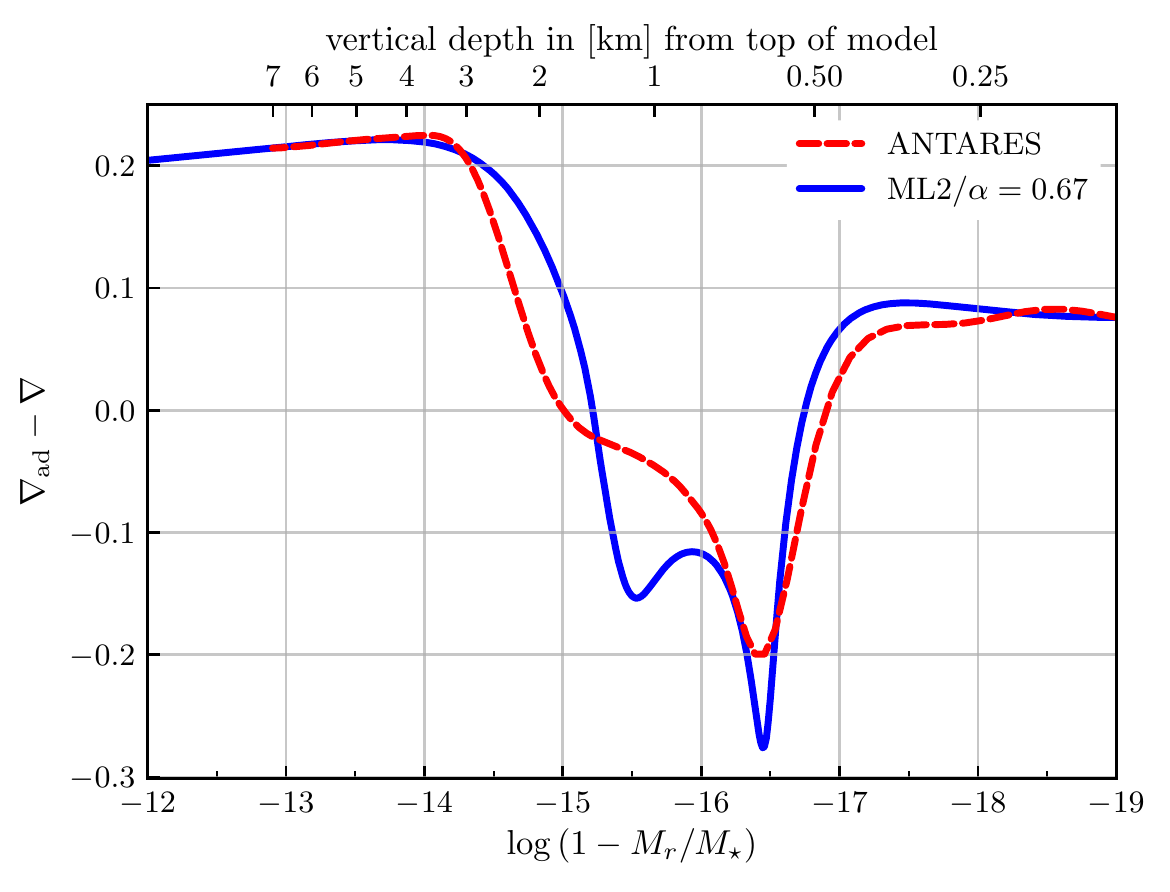}
    \caption{Adiabatic minus average temperature gradient from the numerical simulation
       (ANTARES) and a 1D model based on the ML2 mixing length
       model with an $\alpha$--parameter of $0.67$. 
       }
    \label{fig:tempgradient}
\end{figure}

In Fig.~\ref{fig:tempgradient} we compare $\nabla_{\rm ad} - \nabla$ of the simulation
averages with results from a 1D stellar model with ML2 convection
model \citep{bohm71} using $\alpha=0.67$, where $\alpha$ has been
adjusted to provide the best fit to the final state of the 3D
simulation. We observe that the simulation averages and the 1D model
merge towards the interior, below layers where
$\log(1-M_r/M_{\star}) \gtrsim -13.5$. Above that layer differences in
the gradients occur due to overshooting modifying the stratification,
i.e.\, where $-13.5 \gtrsim \log(1-M_r/M_{\star}) \gtrsim -15$. The
superadiabatic feature at $\log(1-M_r/M_{\star}) \approx -16.4$ is
slightly broader and shallower in the simulation than it is for the
stellar model. Differences further above it are due to the different 
treatment of radiative transfer and the upper boundary conditions 
in the simulation and the 1D model.

% shift log10(column density) by -14.115 to obtain log10(Ms) = log10(1-M_r/M_{\star})

\begin{figure}
	\includegraphics[width=\columnwidth]{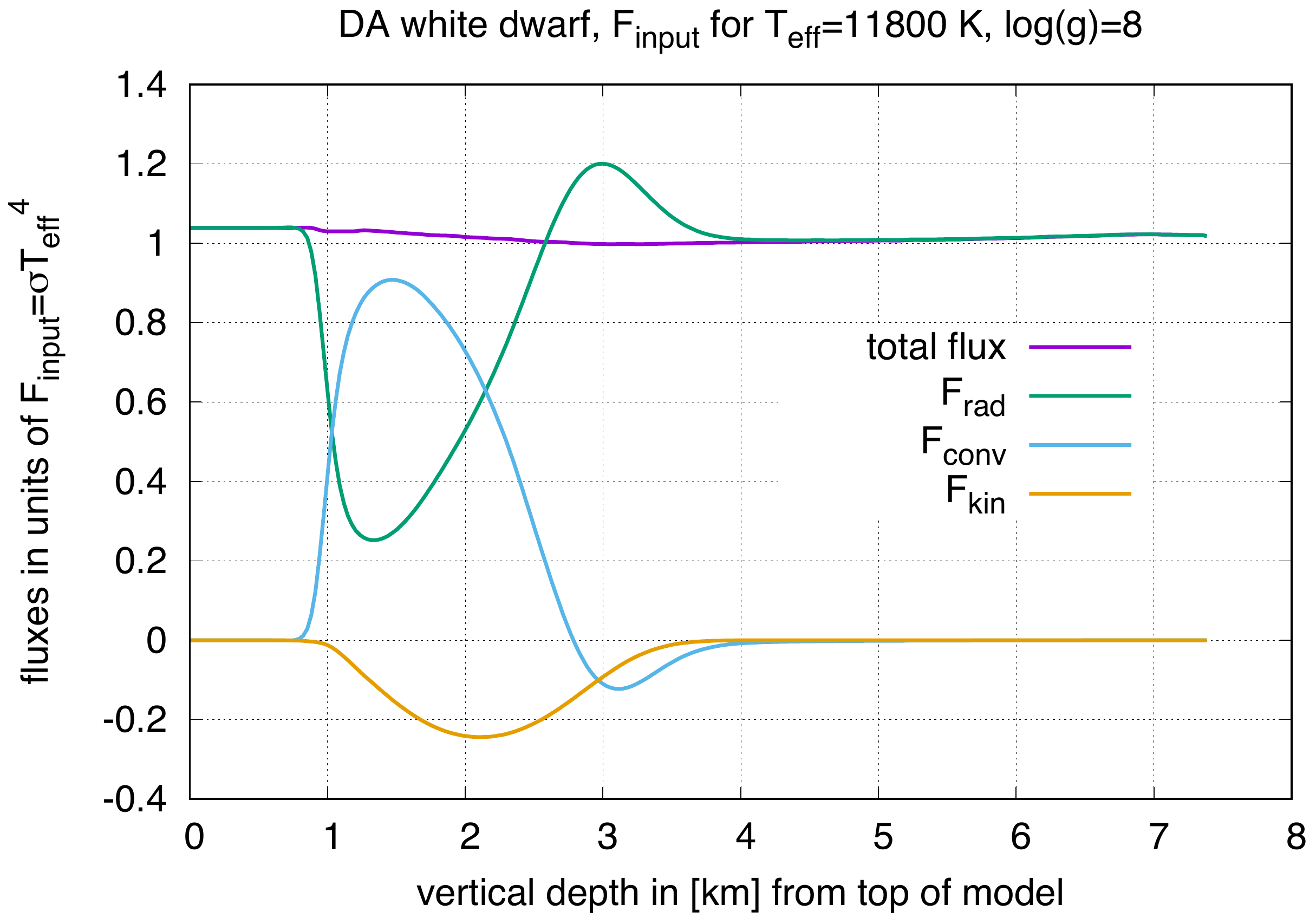}
        \caption{% Horizontal and temporal averages of 
          Radiative, convective, and kinetic energy fluxes, denoted as
          $F_{\rm rad}$, $F_{\rm conv}$, and $F_{\rm kin}$, as well as
          their sum $F_{\rm total}$, each normalised relative
          to the input flux of the 1D starting model, plotted against
          depth measured from the top of the simulation box.}
    \label{fig:fluxes}
\end{figure}

In Fig.~\ref{fig:fluxes} we compare the contributions of radiative, convective, and kinetic
energy fluxes in the vertical direction to the total (vertical) energy flux. The Schwarzschild
unstable region ranges from 0.8~km down to 2~km (as measured from the top of the simulation
box). This is followed by a large overshooting region with non-zero convective and kinetic
energy fluxes. The latter essentially vanish around 4~km. 

\subsection{Relaxation}

We discuss in more detail the relaxation process during our 3D simulation and
the accuracy which we can expect for our statistical data averaged over $t_{\rm stat}$.
Already from the total energy flux depicted in Fig.~\ref{fig:fluxes}, which is obtained from
averaging the horizontal average of $F_{\rm total}$ over $t_{\rm stat}$, we can expect
the simulation to be quite close to thermal relaxation, since the deviations of $F_{\rm total}$ from 
$F_{\rm input}$ are about $4\%$ or less. In \cite{kupka16b} it is explained in detail why we
can expect a simulation to require thermal relaxation only for the {\em upper part} of the 
simulation domain, if it is started from an initial condition where the {\em lower part}
of the computational domain --- located more closely to the stellar interior --- is already in 
a (nearly) relaxed state. This is possible if the lower part is either stratified 
quasi-adiabatically as in simulations of solar granulation or radiatively as in the present case, 
since both can easily and sufficiently accurately be guessed from a 1D model. Indeed,
for the solar case fast relaxation from various initial conditions, all quasi-adiabatic for
the solar interior, was demonstrated with ANTARES by \cite{grimm-strele15b}
(see their Fig.~9 and~10).

However, there is a region in the DA white dwarf considered here, for which the thermal structure
cannot be easily guessed from a 1D model: the region of overshooting below the convection
zone. There, $F_{\rm rad}$ is up to 20\% larger than the total flux (see Fig.~\ref{fig:fluxes}).
These layers require thermal relaxation and thus enforce a minimum relaxation time much larger
than in surface convection simulations of stars with a lower boundary placed inside a quasi-adiabatic 
convection zone such as that one of our Sun (cf.\ \citealt{kupka16b}).

\begin{figure}
	\includegraphics[width=\columnwidth]{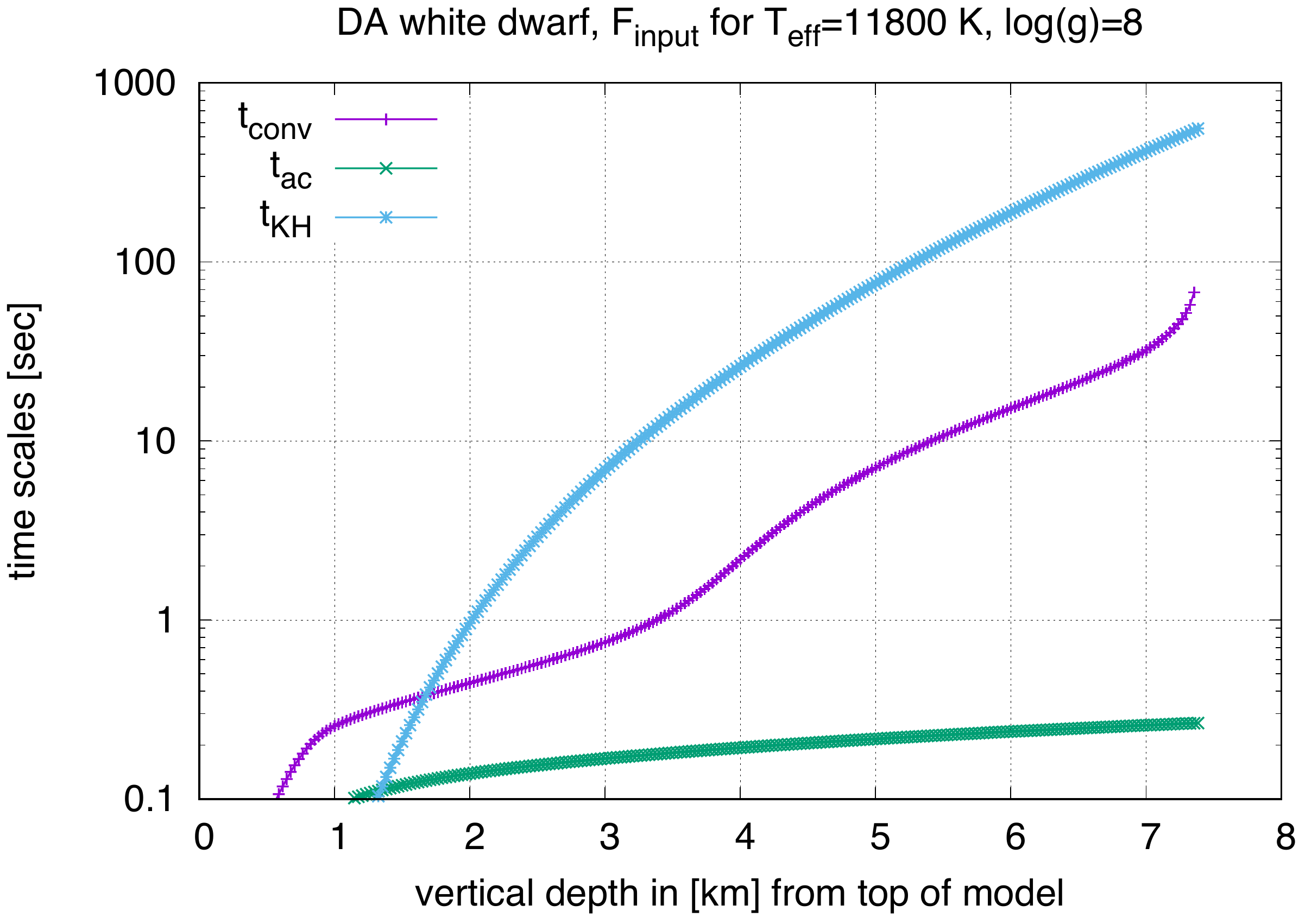}
        \caption{Integral time scales of interest for relaxation computed here as a function
                    of model depth: the convective turnover time $t_{\rm conv}$, acoustic time $t_{\rm ac}$,
                    and Kelvin-Helmholtz time $t_{\rm KH}$.}
    \label{fig:relax-timescales}
\end{figure}

In many cases of astrophysical interest the thermal relaxation time associated with a certain layer of 
a star is approximately equal to the Kelvin-Helmholtz time scale for that layer (cf.\ Chap.~2.3.4 in \citealt{kupka16b}
and also Chap.~5 and~6 in \citealt{kippenhahn94b}, also for discussions of limitations of applicability 
of this approximation): $t_{\rm therm}(x)\approx t_{\rm KH}(x)$. The latter is obtained from integrating 
$t_{\rm KH}(x)\equiv -3 (\int_a^b p \rho^{-1}dM_s)/L_r$, where $a$ is the mass near the surface of the star 
(or the simulation box, here $\approx 0$) and $b$ is the mass contained above that layer at depth 
$x$, i.e., $M_s$. We recall that $p$ and $\rho$ are pressure and density, whereas $L_r=4\pi r^2 F_{\rm total}$ 
is the local luminosity. It is straightforward to adopt this to a box-in-a-star configuration with plane parallel 
geometry and obtain a practical definition for $t_{\rm KH}(x)$, 
namely: $t_{\rm KH}(x)\equiv \langle\, 3 (\int_0^x p(x') dx')/F_{\rm total}\,\rangle_{\rm h,t}$, where the 
integration occurs over vertical location $x'$, with $0 \leqslant x' \leqslant x$, and the result is 
averaged horizontally and in time --- or a horizontally averaged pressure is time averaged, 
as we have done here. In Fig.~\ref{fig:relax-timescales} we plot this calculation of $t_{\rm KH}(x)$ for our 
simulation as a function of depth along with the acoustic time $t_{\rm ac}(x)$, i.e., the time 
a sound wave requires to travel from the top of a simulation box to a particular layer at the vertical point $x$. In
addition, we display the convective time scale $t_{\rm conv}(x)$, which is likewise obtained from integrating the 
inverse of the time and horizontally averaged root mean square of the fluctuating part of the vertical velocity, i.e., 
the square root of $w^2_{\rm rms}=\langle (w-\langle w \rangle_{\rm h})^2\rangle_{\rm h,t}=\langle w'^2\rangle_{\rm h,t}$,
from top down to $x$. Clearly, $\tau_{\rm scrt}=t_{\rm ac}(x_{\rm bottom})$. Now before statistics can be collected
(i.e., ahead of averaging results over the statistical sampling time $t_{\rm stat}$), the time integration has to be first
performed long enough such that initial perturbations both of the thermal mean structure (given by $p$, $T$, $\rho$) 
and the velocity field no longer influence the result and this is just the relaxation time scale $t_{\rm relax}$. Following 
the discussion on thermal relaxation in \cite{kippenhahn94b} and on relaxation of hydrodynamical simulations of stellar
convection in \cite{kupka16b} we can expect $t_{\rm relax}=\max(t_{\rm KH}(x_1),t_{\rm conv}(x_2))$, i.e., the maximum of 
$t_{\rm KH}(x_1)$, evaluated at a layer $x_1$ below which the stratification is essentially in thermal equilibrium already 
from the beginning, and of $t_{\rm conv}(x_2)$, where $x_2$ is evaluated close to the bottom of the simulation domain 
(or at the bottom in case of open boundary conditions). We find $F_{\rm total} \approx F_{\rm rad}$ below $x_1=4$~km 
throughout most of our simulation. Precisely, their difference drops strictly monotonically from $\sim 0.8\%$ at 4~km to 
$\lesssim 0.1\%$ at 5.5~km, and $\lesssim 0.01\%$ at 7~km. In the same region, $F_{\rm total}$ differs from 
$F_{\rm input}$ on average by $\lesssim 1.5\%$ and at most by $< 2.3\%$. We choose $x_2$ slightly above 
$x_{\rm bottom}$ to avoid the layers where $w^2_{\rm rms} \rightarrow 0$. We conclude that 
$t_{\rm KH}(x_1) \approx 25~{\rm s} \approx 106\, \tau_{\rm scrt}$ and 
$t_{\rm conv}(x_2) \approx t_{\rm conv}(x=7\,{\rm km}) \approx 30~{\rm s} \approx 127\, \tau_{\rm scrt}$. So we 
expect the relaxation of the thermal stratification and the velocity field to require up to roughly $130\, \tau_{\rm scrt}$.

\begin{figure}
	\includegraphics[width=\columnwidth]{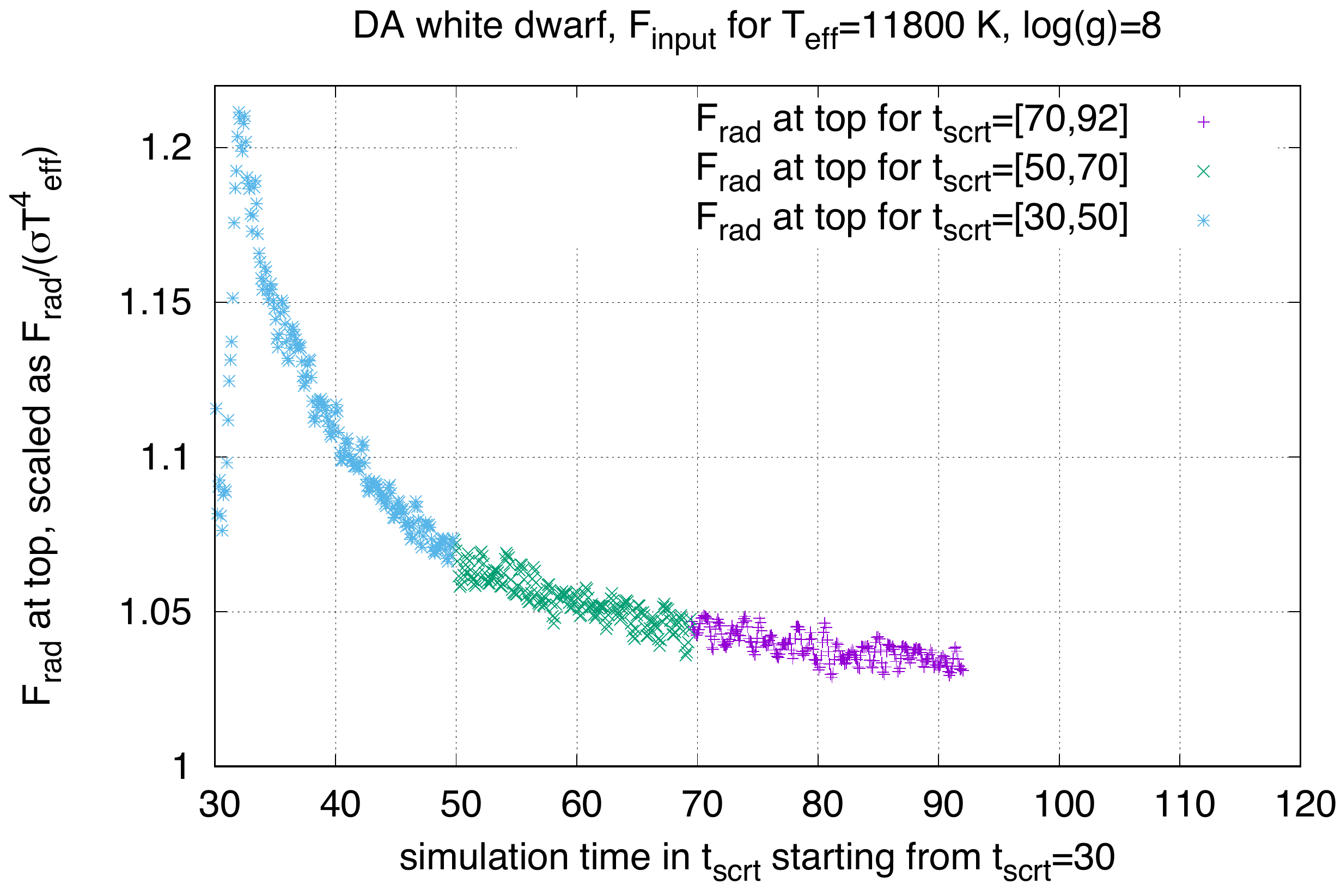}
        \caption{Thermal relaxation is shown here by the time evolution of the emerging radiative flux 
                    at the top of the simulation, normalised by the input flux, plotted as a function of model age 
                    $t_{\rm scrt}$, and grouped into three consecutive sets of data.}
    \label{fig:relax-fluxes}
\end{figure}

To show the degree of thermal relaxation and the simulation time required to achieve it we evaluate
$F_{\rm rad}$ at the top of the simulation box to plot it against model age measured in $t_{\rm scrt} = t/\tau_{\rm scrt}$
where $t$ is the model age in seconds. We note that until $t_{\rm scrt} = 30$ the simulation has been
evolved in 2D and has then been reset through horizontally averaging that state to provide the initial condition 
of the 3D simulation, as described in Sect.~\ref{sec:numsim}. For a thermally relaxed model we expect
$F_{\rm rad}(x_{\rm top}) \approx F_{\rm input}$. In Fig.~\ref{fig:relax-fluxes} we see the result for
our simulation from the beginning of the 3D setting. During the first 20~scrt there is obviously some major 
readjustment going on until a convergence of $F_{\rm rad}(x_{\rm top}) \rightarrow F_{\rm input}$ sets
in which is roughly proportional to ${t_{\rm scrt}}^{-1/2}$ during the time over which we have performed
the simulation (halving the flux difference with respect to $F_{\rm input}$ requires to continue
the simulation for twice the amount of time). We have decided to also drop the next segment of 20~scrt 
for statistical evaluation and begin with the computation of $t_{\rm stat}$ at $t_{\rm scrt} \approx 70$. 

Averaged over the time interval of $t_{\rm stat}$, we find an emerging radiative flux 
$F_{\rm rad}(x_{\rm top}) / F_{\rm input} \approx 1.03845$, when normalizing it relative to $F_{\rm input}$.
This corresponds to a $T_{\rm eff}$ of 11912~K and may be compared with an equivalent of 12197~K 
obtained when averaging over the {\em first} $\rm 10\,scrt$ of the 3D simulation, at the beginning of relaxation. 
For the average over the {\em last} $\rm 5\,scrt$ of $t_{\rm stat}$ we find $F_{\rm rad}(x_{\rm top}) / F_{\rm input}$ 
to have dropped already below $1.035$, whence $T_{\rm eff} \approx 11900$~K. From the time dependence
seen in Fig.~\ref{fig:relax-fluxes} we expect that a doubling of simulation time from $t_{\rm scrt} = 30$
onwards, i.e., up to $t_{\rm scrt} \approx 154$, would allow halving the residual flux difference to less
than $2\%$ and thus $T_{\rm eff} \lesssim 11850$~K, in agreement with the relaxation estimate discussed
above for $t_{\rm KH}$ (see Fig.~\ref{fig:relax-timescales}). We note that from an observer's point of
view we might use the emitted surface flux, i.e.\ a $T_{\rm eff}$ of 11912~K, to characterize the simulation
when averaging it from a $t_{\rm scrt}$ of 70 to 92. However, for the present discussion we prefer
to use the input flux at the bottom (equivalent to $T_{\rm eff}=11800$~K) for scaling the results,
since this value is fixed throughout the simulation.

The differences between surface and input radiative flux which are expected to remain at that value 
of $t_{\rm scrt}$ are partially caused by the initial stratification of the lower region not being in perfect 
thermal equilibrium, in particular with respect to its interaction with layers further above. But there are 
also systematic differences introduced by numerical inaccuracies of the radiative transfer solver 
(cf.\ the little in dip in $F_{\rm total}$ around a depth of 1~km in Fig.~\ref{fig:fluxes} --- we point out 
here that this quantity is not often shown in publications on numerical simulations of stellar surface 
convection, but when it is, similar features are found for layers near the stellar surface). Other sources
of systematic differences of similar order or smaller occur when comparing grey with 
non-grey radiative transfer (cf.\ \citealt{grimm-strele15b} and also \citealt{tremblay11b}), or 
they originate from the detailed implementation of the input boundary condition as well as
from differences between the numerical approximation and assumed microphysics of the 1D starting 
model in comparison with the numerical simulation. Although one could try to ``remove'' residual flux 
differences originating from these sources by longer relaxation, it makes neither physical nor mathematical sense to do so: the 
systematic and numerical errors they contribute to are already roughly equal to the flux difference caused by 
an ``incomplete relaxation'' of the order of 2\% of the total flux.
For the same reason it is also not important whether we estimate $t_{\rm KH}(x)$ by averaging over
$t_{\rm stat}$, as we have done, or calculate it from the initial condition to guide the simulation.
An estimate of the numerical errors due to resolution on the mean temperature profile as obtained from numerical 
simulations of solar granulation with ANTARES has been made by \cite{grimm-strele15c}: the accumulated error
over several sound crossing times --- for a {\em relative} resolution roughly comparable to the one used in the present
work --- was found in the range of a fraction of one percent, with maximum errors up to a few percent. The error
calculated that way is chiefly due to finite resolution in the numerical simulation and since the relative 
resolutions, maximum Mach numbers, etc., are {\em roughly} comparable to our present case, we should expect
relative (numerical) errors of similar size also for the present simulation. 
Moreover, from Fig.~\ref{fig:relax-fluxes} we can see the impact of oscillations present in the simulations:
they lead to a variation in $F_{\rm rad}(x_{\rm top}) / F_{\rm input}$ of $\pm 0.5\%$ over $\tau_{\rm scrt}$
(we discuss those further below). 

\begin{figure}
	\includegraphics[width=\columnwidth]{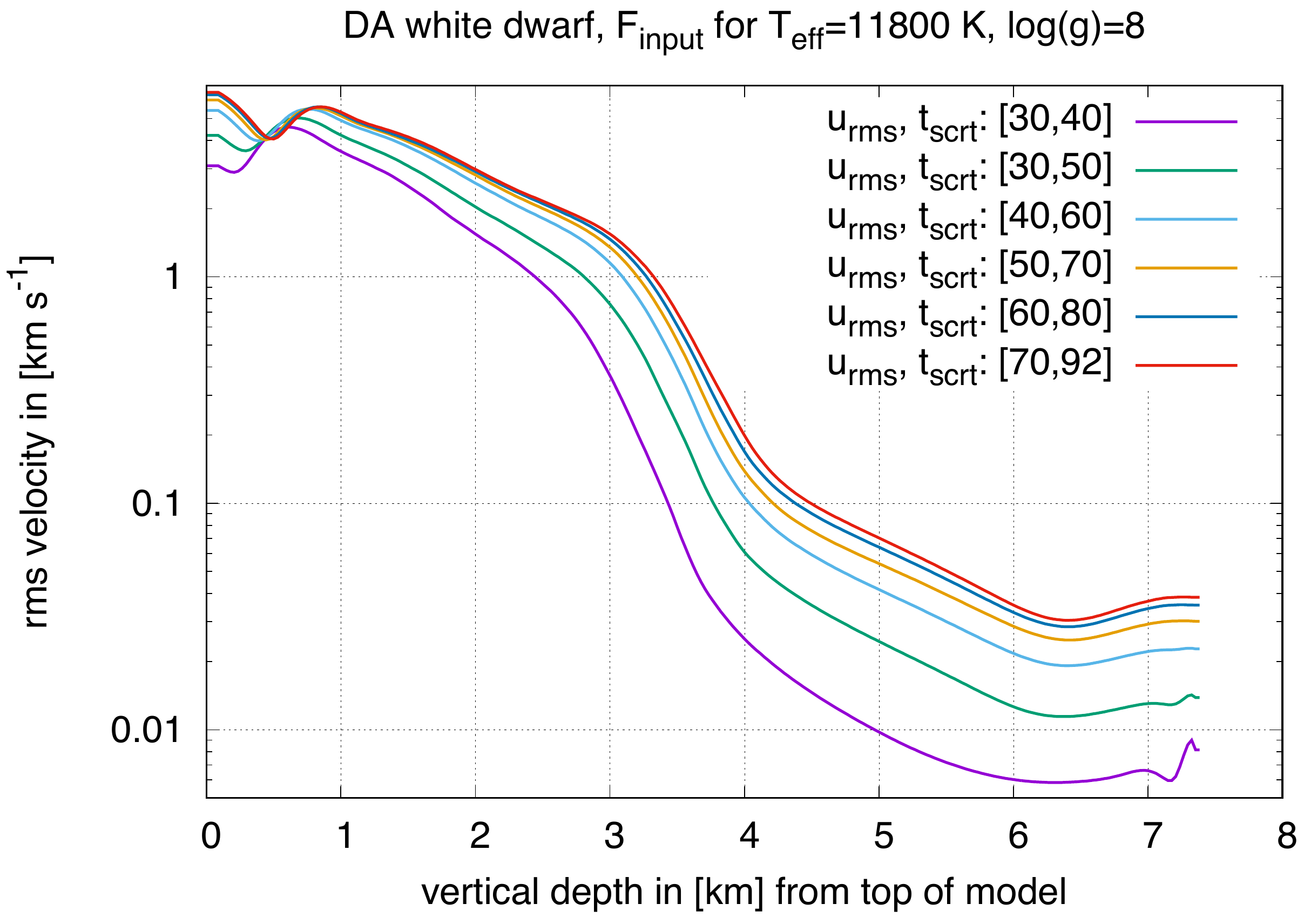}
        \caption{Relaxation of the root mean square average of the fluctuation of the $u$-component of 
                     horizontal velocity around its horizontal mean, $u_{\rm rms}$, displayed as a function of depth and 
                     time averaged over a sequence of intervals distinguished by model age $t_{\rm scrt}$.}
    \label{fig:relax-urms}
\end{figure}

It is thus of little practical use to extend the present simulation over a longer interval in time.
If at all, one may consider $t_{\rm scrt} \rightarrow 150$, since beyond this value the systematic
errors become dominant. But as we show here, also for 
sufficiently accurate statistics of the velocity field time integration beyond $t_{\rm scrt} \approx 92$ 
is not necessary. Fig.~\ref{fig:relax-urms} provides an example for the rapid convergence of the velocity field 
towards a statistically stationary state well within our estimate of $t_{\rm conv}(x_2)$ discussed along
Fig.~\ref{fig:relax-timescales}. We find halving of the relative error of the horizontally and time averaged root mean 
square of the fluctuation of the $u$-component of horizontal velocity, $u_{\rm rms}$,  relative to its instantaneous horizontal 
mean for each consecutive interval of 10~scrt, i.e., $\propto {t_{\rm scrt}}^{-2}$. As can be estimated from the mere 
$12\%$ increase through most of the layers of the simulation when progressing from a $t_{\rm scrt}$ in the interval 
$[60,80]$ to $[70,92]$, i.e.\ the range of $t_{\rm stat}$, quadratic convergence is found for nearly the entire simulation 
duration of $t_{\rm scrt}$. This is fast enough such that the difference between these two averagings is 
already the entire expected growth if the computation were continued beyond a $t_{\rm scrt}$ 
of about 150. Fig.~\ref{fig:relax-urms} thus also provides an error estimate for $u_{\rm rms}$ with respect
to its relaxation.

For the surface layers this error is even much smaller and clearly it converges rapidly throughout the 
entire simulation box. This similarly holds for the siblings of $u_{\rm rms}$, i.e., $v_{\rm rms}$ and $w_{\rm rms}$, 
which characterise the second horizontal and the vertical component of the flow. Also skewness and
kurtosis of velocity and temperature fields, i.e., higher-order statistical correlations that we discuss below
in Sect.~\ref{subsec:skew_kurt}, converge fast but for a limited region dominated by
single events which is explained there in more detail. We can hence safely use velocity and temperature
statistics averaged over $t_{\rm scrt}$ in the following discussion. From a physical 
point of view this fast convergence of the statistics for the velocity field irrespectively of the less accurate 
convergence of the emerging radiative flux is not surprising: the velocity fields are generated by convective 
processes occurring in the upper part of the simulation box (with very short relaxation times). All 
what is left is a small drift as a function of time, which does not affect the functional form, physical 
processes, and even the level of accuracy we expect from the velocity related quantities that we 
discuss below. The independence of the residual error in the velocity field from the degree of thermal
relaxation is also corroborated by the completely different convergence rates, $\propto {t_{\rm scrt}}^{-2}$
for $u_{\rm rms}$ and $\propto {t_{\rm scrt}}^{-1/2}$ for the total (radiative) flux at the surface. We expect 
the faster convergence rate of velocities to change to the smaller one of thermal relaxation once 
$t_{\rm scrt} \gtrsim t_{\rm conv}(x_2)/ \tau_{\rm scrt} \approx 127$. As discussed in the context of thermal
relaxation, at this point in time evolution the model intrinsic errors dominate over the residual error, and
the accuracy reached for $u_{\rm rms}$ and related quantities at  $t_{\rm scrt} \in [70,92]$ is already adequately
small. Thus, our simulation data obtained over $t_{\rm stat}$ are both based on a sufficiently well relaxed 
simulation and have statistical errors small enough to be useful subsequently.

\subsection{Velocities}   \label{subsec:velocities}

\begin{figure}
	\includegraphics[width=\columnwidth]{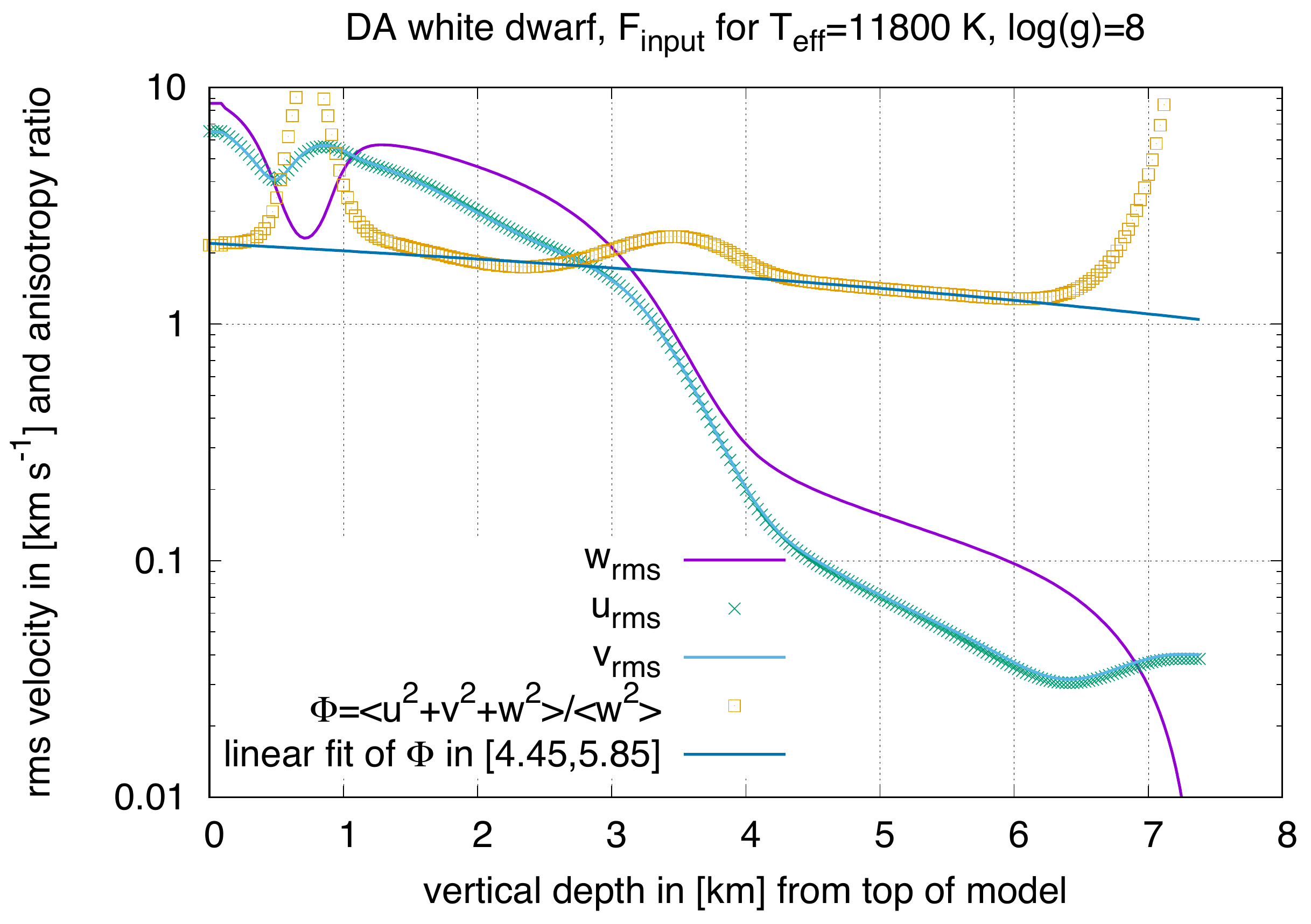}
        \caption{Vertical rms velocity, $w_{\rm rms}$, and its  
        horizontal counterparts, $u_{\rm rms}$ and $v_{\rm rms}$, perfectly agreeing,
        on a logarithmic scale. 
        The dimensionless anisotropy ratio $\Phi$ is also plotted together with its
        linear fit computed for the interval from 4.45~km to 5.85~km.}
    \label{fig:velocities}
\end{figure}

As the vertical and horizontal velocity fields are reasonably well
converged in the convection zone and in the overshooting region
underneath, we can analyse the time and horizontally averaged root
mean square of their fluctuating components, i.e., $w_{\rm rms}$ 
for the vertical velocity, and likewise $u_{\rm rms}$ and $v_{\rm rms}$ 
for the two horizontal components. In Fig.~\ref{fig:velocities} we plot them
on a logarithmic scale. From local maxima of $\sim 5.7~{\rm km\, s}^{-1}$ 
near the top of the convective zone they gradually drop towards its bottom. 
Where $F_{\rm conv}$ changes sign (see Fig.~\ref{fig:fluxes}), $u_{\rm rms}$ 
and $v_{\rm rms}$ have reached $\sim 1.8~{\rm km\, s}^{-1}$ while $w_{\rm rms}$ 
is still $\sim 2.8~{\rm km\, s}^{-1}$, but has begun to decay much faster, a process 
occurring for $u_{\rm rms}$ and $v_{\rm rms}$ again from around 3.2~km onwards. 
The horizontal velocity components begin to dominate the total kinetic energy and 
$\Phi = (w_{\rm rms}^2+u_{\rm rms}^2+ v_{\rm rms}^2)/w_{\rm rms}^2$
exceeds a value of 2 with a local maximum around 3.5~km. 
There, a rapid, exponential decay sets in. It is slightly larger for 
$u_{\rm rms}$ and $v_{\rm rms}$. For $w_{\rm rms}$ the decay slows 
down around 4~km, where $F_{\rm conv}$ begins to vanish
($-1\% \lesssim\, F_{\rm conv}/F_{\rm input} \lesssim\, 0\%$). 
The decay of $u_{\rm rms}$ and $v_{\rm rms}$ continues to be rapid 
down to about 4.2~km and $\Phi < 2$. The simulation can be used
for an accurate study of even deeper layers where the lower  
boundary is still sufficiently away to avoid direct interference. 
We find that velocities decay at a slower rate than in
the ``overshooting zone proper". Interestingly, $u_{\rm rms}$ and
$v_{\rm rms}$ continue to decay more rapidly than $w_{\rm rms}$:
over an extended region from about 4.45~km to 5.85~km $\Phi$ 
features a nearly linear decrease with depth which ceases
only once the influence of the lower (closed) boundary condition
becomes notable. Fig.~\ref{fig:velocities} highlights this relation by
a linear least squares fit of $\Phi$ for that region. Note that the
logarithmic scale in that figure would actually suggest an exponential
fit for $\Phi$ while a plot in linear scale motivates the linear model
function chosen. This small difference is caused by a very low
e-folding scale: in this case an actually exponential function can
be approximated by a linear function over an extended range. We 
return to a comparison between both models in Sect.~\ref{subsubsec:wave-dominated}.

\begin{figure}
	\includegraphics[width=\columnwidth]{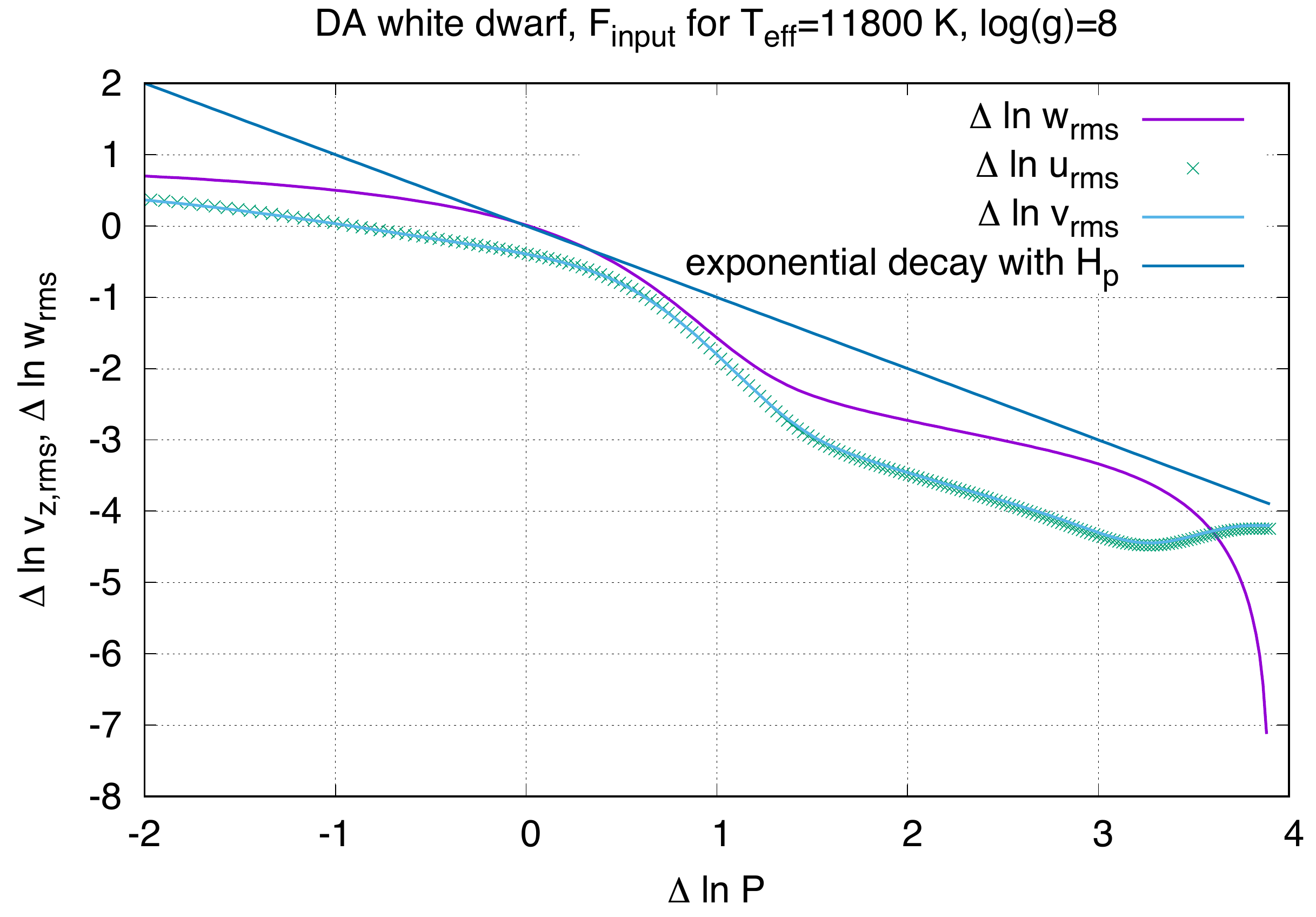}
        \caption{Logarithm of $w_{\rm rms}$, $u_{\rm rms}$, and $v_{\rm rms}$,
          as functions of the logarithm of total pressure, relative to a reference
          velocity and pressure. The exponential decay hypothesis for $w_{\rm rms}$ as 
          function of pressure scale height is indicated.}
    \label{fig:scaling}
\end{figure}

\begin{figure}
	\includegraphics[width=\columnwidth]{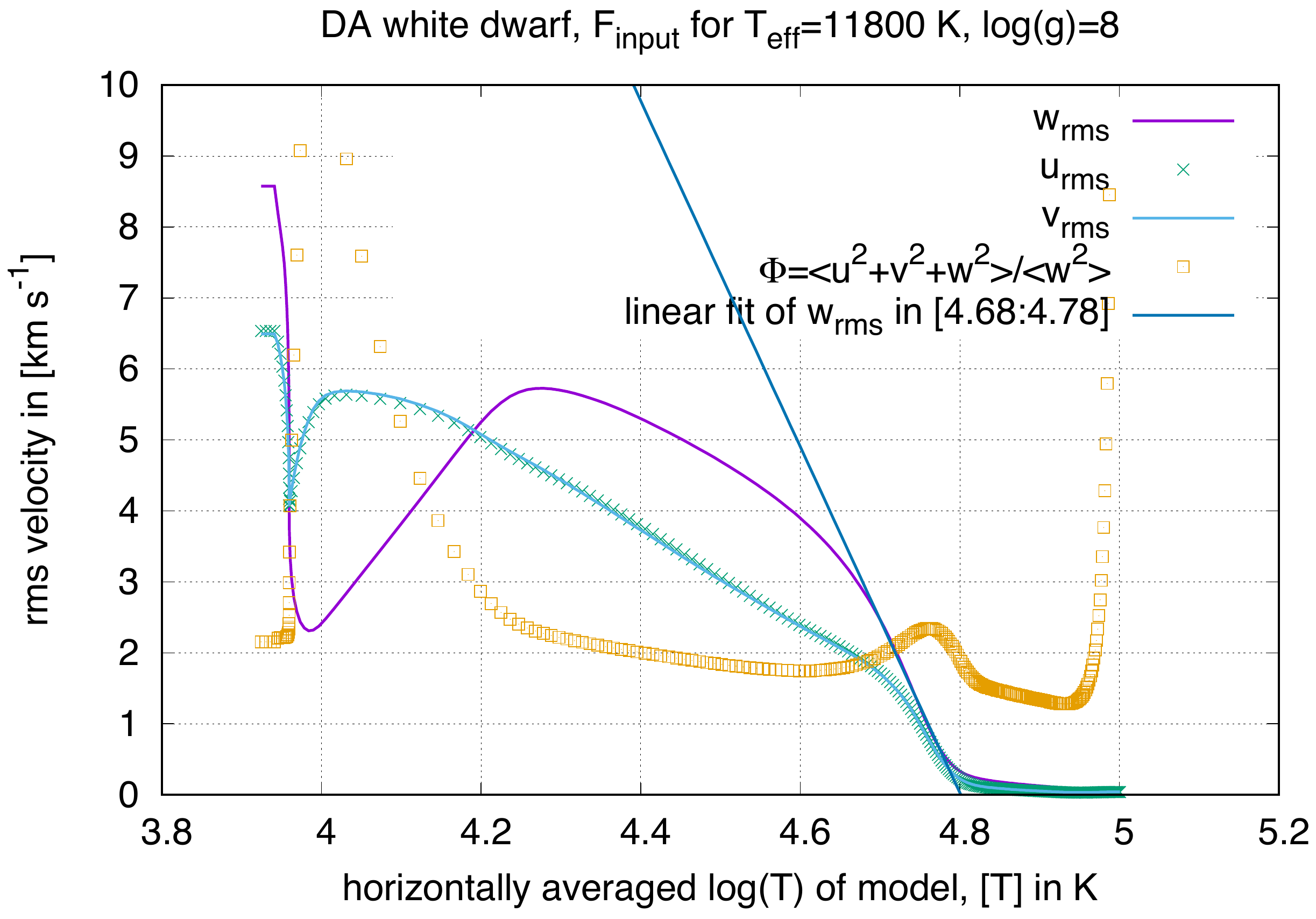}
        \caption{$\Phi$ as well as $w_{\rm rms}$, $u_{\rm rms}$, and $v_{\rm rms}$ 
        as functions of $\log T$ alongside a linear fit of $w_{\rm rms}$ in the 
        overshooting region (see text).}
    \label{fig:logT}
\end{figure}

In Fig.~\ref{fig:scaling} we plot the logarithm of $w_{\rm rms}$ as well as $u_{\rm rms}$ 
and $v_{\rm rms}$ as functions of $\ln\, P$, normalised relative to the velocity and 
the total pressure at the layer where $F_{\rm conv}$ changes sign (at 2.785~km in 
Fig.~\ref{fig:fluxes}). We also plot a line to indicate an exponential decay of the vertical 
velocity field with $H_{\rm p}$, as proposed in \cite{tremblay15b} to occur for DA WDs.
We cannot identify a {\em unique} exponential scaling law for $w_{\rm rms}$ in regions 
where $F_{\rm conv} < 0$. For the layers where $\Delta \ln P$ is between 0 and 1.5 
($|F_{\rm conv}/F_{\rm input}| > 1\%$), the decay rate is first about $1\, H_{\rm p}$, then 
becomes twice as steep $(0.5\, H_{\rm p})$, then settles at a quarter of that rate
$(2\, H_{\rm p})$. Similar holds for $u_{\rm rms}$ and $v_{\rm rms}$ with a shift 
in location and different decay rates particularly for $\Delta \ln P \gtrsim 1.5$. 
No simple polynomial law can describe this dependency either (\citealt{canuto97d}
derived a polynomial decay as function of distance from the convection zone if
the dissipation rate of turbulent kinetic energy were computed from its {\em local limit} 
expression, cf.\ Fig.~\ref{fig:velocities}).

In turn, $w_{\rm rms}$ depends linearly on $\log T$ from 2.73~km, 
where $F_{\rm conv} \gtrsim 0$, down to about 3.73~km, where 
$F_{\rm conv}/F_{\rm input} \approx -2\%$ (see Fig.~\ref{fig:logT}). 
The linear fit of $w_{\rm rms}$ as a function of $\log T$ finds 
$w_{\rm rms}$ to become zero where $F_{\rm conv}/F_{\rm input} \approx -0.87\%$,
although this occurs below the domain of its validity. There, $\log T \approx 4.80$ 
at a depth of $\approx 3.97$~km. Inside the fit region, $\Delta \ln P$ increases from $-0.07$ 
to $1.04$. A linear decay of $w_{\rm rms}$ as a function of $\log T$ for exactly the same 
part of the overshooting region has been reported by \cite{montgomery04b} for hotter 
DA WDs with $T_{\rm eff} \gtrsim 12200$~K when solving the Reynolds stress model 
of \cite{canuto98b}. Note that the same scaling could also be inferred from Fig.~4 
of \cite{tremblay15b} for their models with a $T_{\rm eff}$ of 12100~K, 12500~K, 
and, roughly, also 13000~K for ``zone~3" (cf.\ their Table~1).
As hydrogen is fully ionised in that zone and $F_{\rm total} \approx F_{\rm rad}$, both
$T$ and $H_{\rm p}$ scale linearly with depth, thus our Fig.~\ref{fig:logT} and Fig.~1 
of \cite{montgomery04b} imply the same {\em (non-) linear decay of $w_{\rm rms}$ with
$T$ and $H_{\rm p}$}, albeit (cf.\ Fig.~\ref{fig:scaling}) {\em this also only holds for 
a limited region}.

\section{Analysis}    \label{sec:analysis}

\begin{figure*}
\centering

    \includegraphics[width=0.68\columnwidth]{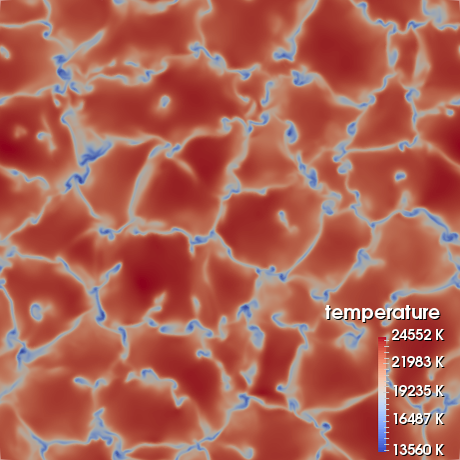}  \hspace{1.3mm}% left column, first row
    \includegraphics[width=0.68\columnwidth]{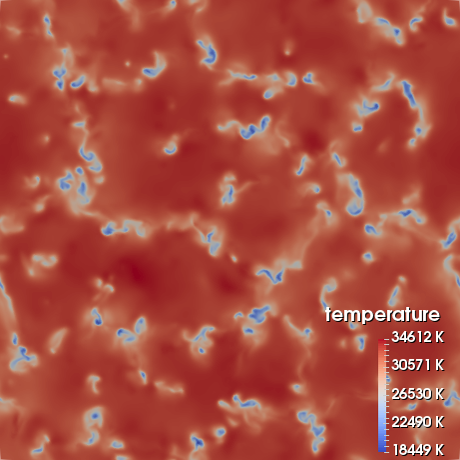} \hspace{1.3mm}% right column, first row
    \includegraphics[width=0.68\columnwidth]{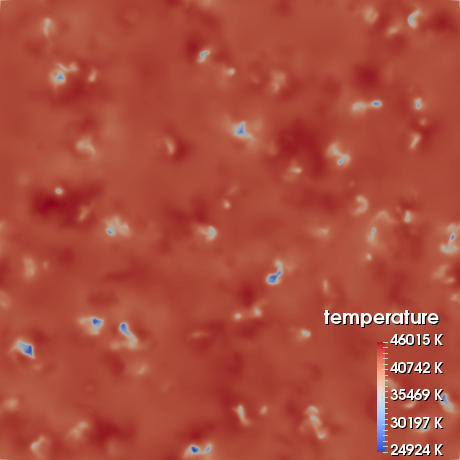} \\
    \vspace{1.8mm}% left column, second row

    \includegraphics[width=0.68\columnwidth]{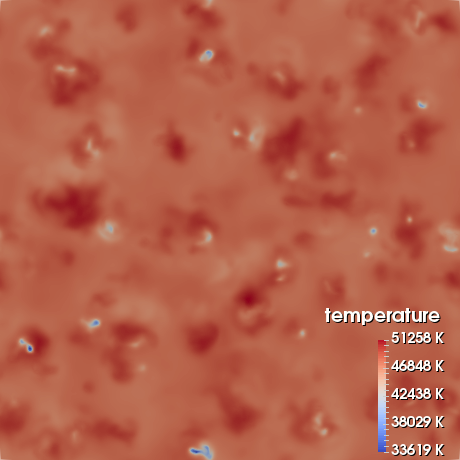} \hspace{1.3mm}% right column, second row
    \includegraphics[width=0.68\columnwidth]{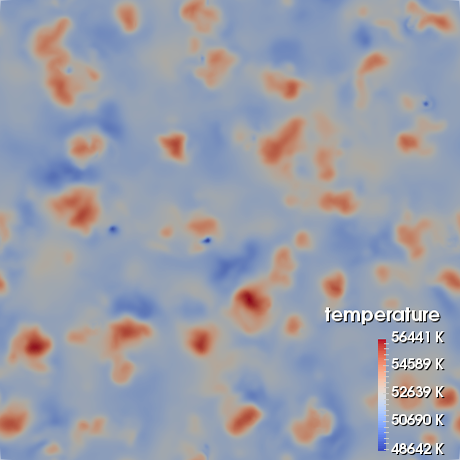} \hspace{1.3mm}% left column, third row
    \includegraphics[width=0.68\columnwidth]{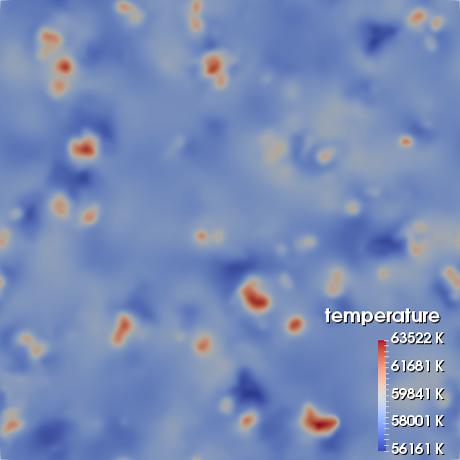} \\
    \vspace{1.8mm}% right column, third row

    \includegraphics[width=0.68\columnwidth]{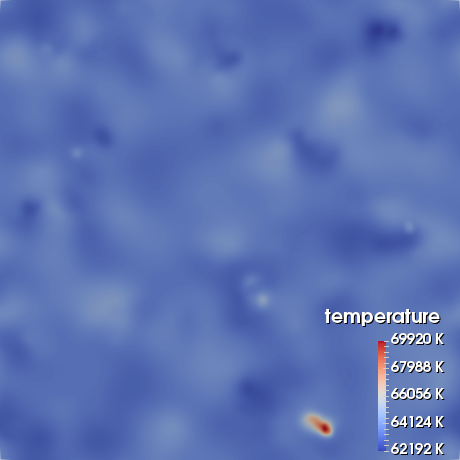} \hspace{1.3mm}% left column, fourth row
    \includegraphics[width=0.68\columnwidth]{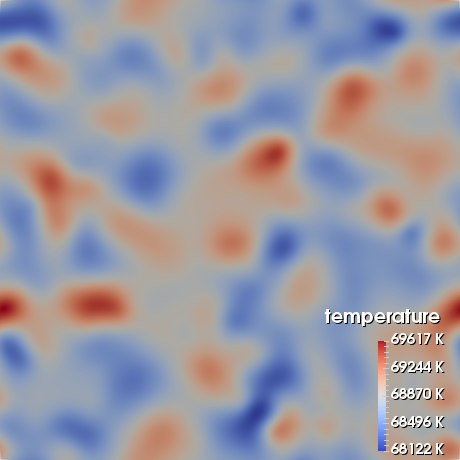} \hspace{1.3mm}% right column, fourth row
    \includegraphics[width=0.68\columnwidth]{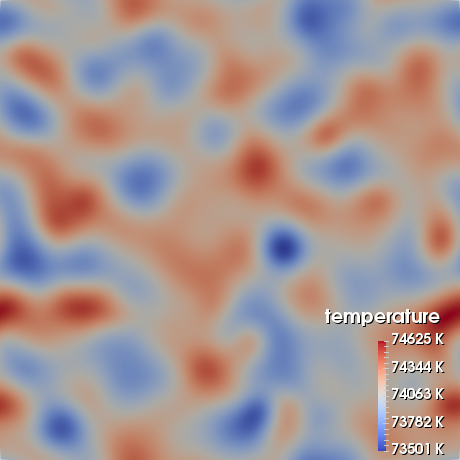} % right column, fourth row

    \caption{Horizontal cuts of temperature as a function of depth at time $t_{\rm scrt} = 92$. 
                 The top row displays snapshots at depths of 1.50~km, 2.00~km and 2.50~km
                 (Fig.~\ref{fig:T_slices}a--c) as measured from the top of the simulation box. The middle row continues 
                 this series for depths of 2.75~km, 3.00~km and 3.50~km (Fig.~\ref{fig:T_slices}d--f). The bottom row displays 
                 the cuts for depths of 4.00~km, 4.50~km and 5.00~km (Fig.~\ref{fig:T_slices}g--i). Note that the 
                 temperature scale changes for each snapshot and the ratio between maximum and minimum value peaks
                 near the Schwarzschild stability boundary around 2.00~km. Clearly inside the plume-dominated 
                 region it is already smaller and drops only gently from 3.00~km to 4.00~km. There is again 
                 a drastic drop in scale for layers inside the wave-dominated region.}
      \label{fig:T_slices}
\end{figure*}

\subsection{The temperature field and overshooting}
 
One may characterise an overshooting zone with respect 
to the sign of the convective flux, the superadiabatic gradient (or alternatively 
the gradients of entropy or potential temperature), the flux of (turbulent) kinetic 
energy, the mass flux, or the local maxima and minima of these quantities.
Assuming a nearly adiabatic temperature gradient for cases of convection
at low radiative loss rates (high Peclet number) \cite{zahn91b} distinguished
a separation of the Schwarzschild unstable convection zone proper ($F_{\rm conv} > 0$, 
$\nabla > \nabla_{\rm ad}$, with  $|\nabla - \nabla_{\rm ad}|/\nabla_{\rm ad} \ll 1$),
from the zone of {\em subadiabatic penetration} ($F_{\rm conv} < 0$, $\nabla < \nabla_{\rm ad}$, 
while again $|\nabla - \nabla_{\rm ad}|/\nabla_{\rm ad} \ll 1$), and the thermal boundary layer 
($F_{\rm conv} < 0$, $\nabla < \nabla_{\rm ad}$,  but $|\nabla - \nabla_{\rm ad}|/\nabla_{\rm ad} \rightarrow O(1)$),
until eventually $F_{\rm conv} \approx 0$ (stable, radiative region, no longer mixed by convection).
This case of penetrative convection he distinguished from the case of {\em overshooting}
for which a more narrow definition was suggested: there, large radiative losses prevent altering the 
temperature gradient towards the adiabatic one in regions where $F_{\rm conv} < 0$.

For the DA white dwarf we consider in the present case, it follows
from Fig.~\ref{fig:fluxes} that radiative losses everywhere in the overshooting 
region are large (low Peclet number). So it makes sense to distinguish, as
in \cite{freytag96b}, the following regions. First of all, the convection zone proper 
with its Schwarzschild boundary ($F_{\rm conv} > 0$, $\nabla > \nabla_{\rm ad}$) and
then an overshooting region where plumes gradually heat up that do not yet experience 
negative net buoyancy (and which we call {\em countergradient region}, 
$F_{\rm conv} > 0$, $\nabla < \nabla_{\rm ad}$). The layer at which the change of sign 
in $F_{\rm conv}$ occurs is termed ``flux boundary'' by some authors,
for example, in \cite{tremblay15b}. From there onwards throughout where 
$F_{\rm conv} < 0$ and $\nabla < \nabla_{\rm ad}$, plumes penetrate into a {\em region of 
counterbuoyancy}. We call this the {\em plume-dominated region} in the following.
One might also consider the local minimum of $F_{\rm conv}$ to mark
the region where the thermal boundary layer begins, as in \cite{zahn91b},
although this has a less important role, as we see in the following figures. 
Finally, motion can persist beyond where $F_{\rm conv} \rightarrow 0$ and 
$\nabla < \nabla_{\rm ad}$, which we further below call the {\em wave-dominated region}. 
The reasons for this naming become evident from the following figures.

In Fig.~\ref{fig:T_slices}a, well inside the convection zone, the network of
intergranular lanes is still clearly visible, similar to the stellar surface
depicted in Fig.~\ref{fig:granules}, although the temperature contrast between
maximum and minimum values has become larger and more skewed (low temperature
regions cover a smaller area). At the Schwarzschild boundary, Fig.~\ref{fig:T_slices}b,
the granular structure is still recognizable, but the intergranular ``network'' of downflows
is no longer connected; rather, it is characterised by plumes, chiefly grouped together near
corner points where several granules meet each other in layers inside the convection
zone. The now isolated, cold downflow columns maintain their identity (Fig.~\ref{fig:T_slices}c), 
although they gradually lose the contrast between them and their environment, until the
``flux boundary'' is reached, where $F_{\rm conv}$ changes sign (Fig.~\ref{fig:T_slices}d).
Below this level the structure rapidly changes: the temperature contrast is
drastically reduced over a small distance, from $3:2$ at the flux boundary around 2.75~km 
to just $7:6$ at 3.00~km (Fig.~\ref{fig:T_slices}e). Now the plumes have
reverted their role: they are hotter than their environment. This structure is maintained
throughout the entire plume-dominated region (Fig.~\ref{fig:T_slices}e--g). The contrast
drops only gently (down to $9:8$), but more importantly, the number of plumes per
area and the area each of them covers rapidly decrease. Indeed, at 4.00~km we
see just one strong plume left in comparison with some three dozen at 3.00~km.
As we know through Fig.~\ref{fig:fluxes}, around 4.00~km, $F_{\rm conv}/F_{\rm total} \rightarrow 0$
(essentially), and a new pattern becomes visible. This pattern remains throughout
the wave-dominated region depicted in Fig.~\ref{fig:T_slices}h--i, with the small contrast between
maximum and minimum temperature slowly decreasing from 2.2\% to 1.5\%. This continues
further into the wave--dominated region (e.g., to a contrast of 1.2\% at 5.50~km, not shown
in Fig.~\ref{fig:T_slices}), although the visible structures have no longer the trivial vertical 
correlation which is easily found for the layers further above. We remark here that the
flow patterns --- which allow an easy distinction between the convectively unstable,
the counter--gradient, the plume--dominated and the wave--dominated region --- are present 
already at $t_{\rm scrt}=35$. They are not related to the patterns of initial relaxation,
say at $t_{\rm scrt}=30.1$, where the wave--dominated region is still mostly
unperturbed ($\Delta T < 1\,{\rm K}$ at 5.00~km) and the pattern visible in
the plume--dominated region has very little, if any, relation to what can be
seen in the wave--dominated region 5~scrt later. We conclude that the patterns
observed in Fig.~\ref{fig:T_slices} are intrinsic to the flow, but not to the initial perturbation 
chosen to start it, either in the wave--dominated region, or in any layer further above.
We skip a similarly detailed discussion of the velocity field and instead demonstrate how
skewness and kurtosis capture many of the statistical and topological properties of
the velocity and temperature field.

\subsection{Skewness and kurtosis of velocity fields and temperature}  \label{subsec:skew_kurt}

\begin{figure}
	\includegraphics[width=\columnwidth]{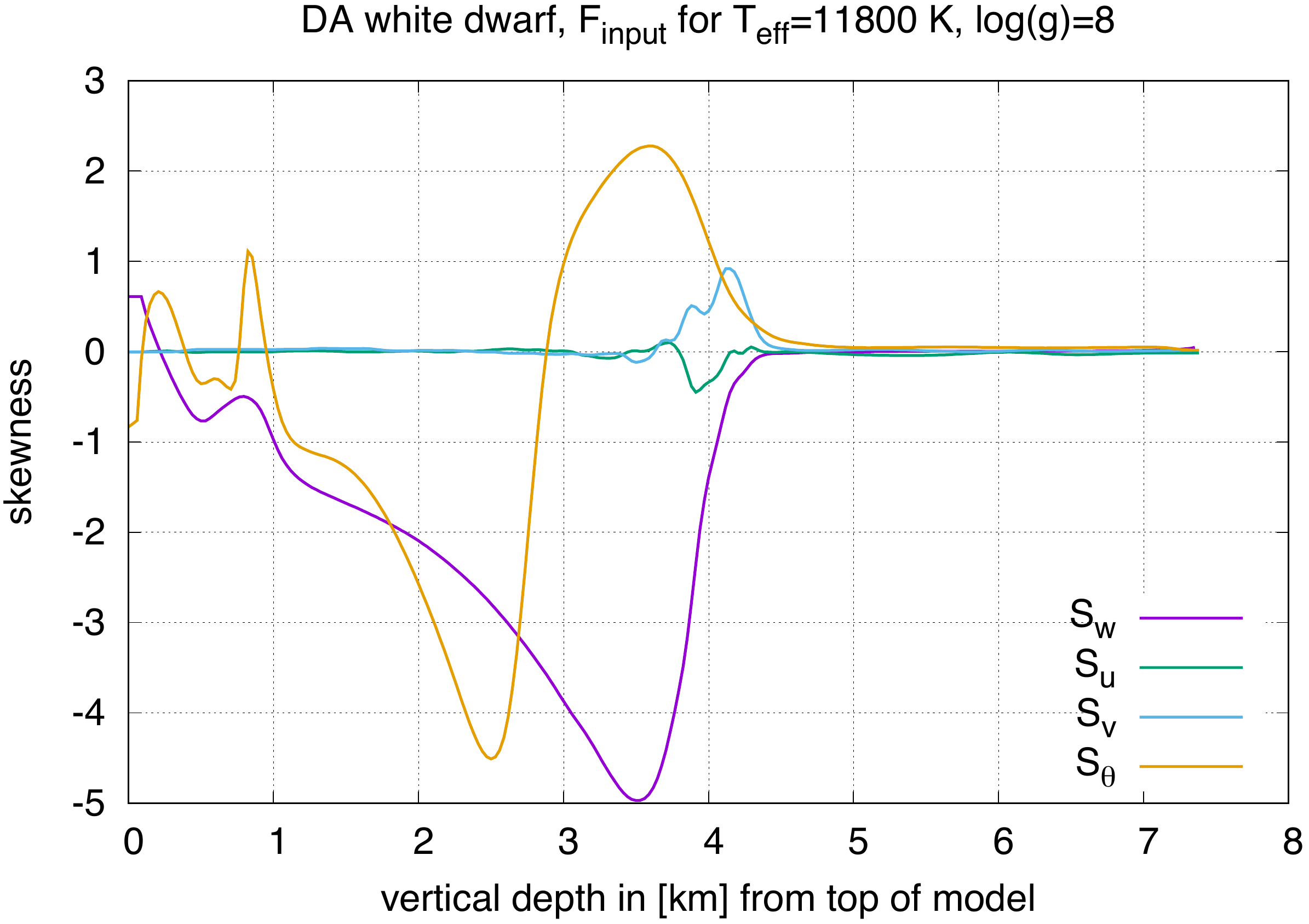}
        \caption{Skewness of fluctuations of the temperature field, $S_{\theta}$, and the
                     components of the velocity field, $S_{w}, S_{u}, S_{v}$, around their horizontal mean, averaged over 
                     each layer and in time (over $t_{\rm stat}$), plotted as a function of depth.}
    \label{fig:skewness}
\end{figure}

In Fig.~\ref{fig:skewness} we display the skewness of fluctuations of the temperature field, $S_{\theta}$, and 
the vertical ($w$) and horizontal components of the velocity field, $S_{w}$, $S_{u}$, and $S_{v}$, around their 
horizontal mean, averaged over each layer and in time (over $t_{\rm stat}$), plotted as a function of depth.
The negative values of $S_{\theta} = \langle (T-\langle T \rangle_{\rm h})^3\rangle_{\rm h,t} / (\langle (T-\langle T \rangle_{\rm h})^2\rangle_{\rm h,t})^{3/2}
=\langle T'^3\rangle_{\rm h,t} / (\langle T'^2\rangle_{\rm h,t} )^{3/2}$ within the convectively unstable zone
indicate that locally low values of $T$ cover a smaller surface area and hence have to deviate further 
from the horizontal mean value than locally high values, in perfect agreement with Fig.~\ref{fig:T_slices}a.
Within the countergradient region this becomes even more pronounced due to the fast columns of downflows generated 
by the strongest remnants of the downflow network (cf.\ Fig.~\ref{fig:T_slices}c). The situation becomes reversed 
around the flux boundary (Fig.~\ref{fig:T_slices}d) and the plumes, which cover a smaller area than the upflow, are now 
hotter than their environment (Fig.~\ref{fig:T_slices}e--f). Hence, $S_{\theta} > 0$ in that region with a pronounced
global maximum where $S_w$ has its global minimum. Eventually, in the wave-dominated region, $S_{\theta} \rightarrow 0$
which demonstrates that the temperature distribution is symmetric around its horizontal mean in those layers.
In comparison, $S_w$, which is defined analogously to $S_{\theta}$, but for the vertical velocity field,
also demonstrates that the downflows cover a more narrow area than the upflows (due to
conservation of mass and momentum), whence $S_w < 0$ in the convection zone. The distribution
becomes more and more skewed once the network of downflows has transformed into a set of plumes
where only the fastest and hottest one can penetrate sufficiently deep (cf.\ Fig.~\ref{fig:T_slices}f and note
its location close to the global extrema of $S_w$ and $S_{\theta}$). Once the plumes disappear in number,
$S_w \rightarrow 0$ and remains near that value for the entire wave-dominated region. Since the simulation
has (or, rather, should have) no preferred symmetry of velocities into any of the horizontal directions, 
we expect $S_u \approx 0$ and $S_v \approx 0$ throughout the simulation and this is also an indicator
for the degree of convergence of the statistical results. We find this confirmed by Fig.~\ref{fig:skewness} with
one exception: around 4~km, both $S_u$ and $S_v$ deviate from $0$ and this can be understood as resulting 
from the statistics depending on very few events with large impact, i.e., on a few plumes managing to penetrate 
deep enough (Fig.~\ref{fig:T_slices}g), leading eventually to sidewards flow by conservation of mass. Many 
more such events are required to obtain a converged statistical result within $t_{\rm stat}$ in that region
at the given horizontal extent of our simulation. 

\begin{figure}
	\includegraphics[width=\columnwidth]{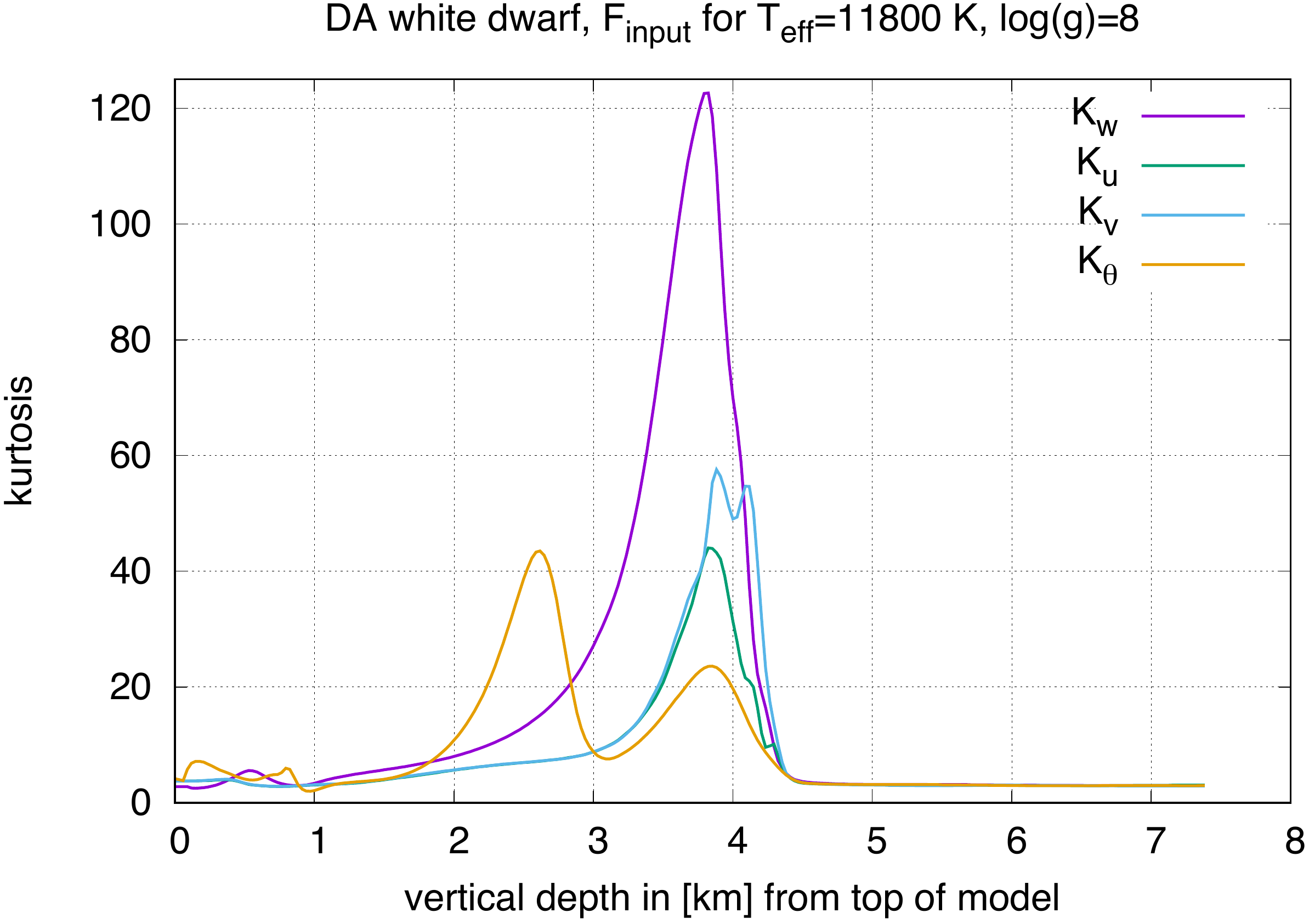}
        \caption{Kurtosis of fluctuations of the temperature field, $K_{\theta}$, and the
                     components of the velocity field, $K_{w}, K_{u}, K_{v}$, around their horizontal mean, averaged over 
                     each layer and in time (over $t_{\rm stat}$), plotted as a function of depth (full range of kurtosis shown).}
    \label{fig:kurtosis}
\end{figure}

\begin{figure}
	\includegraphics[width=\columnwidth]{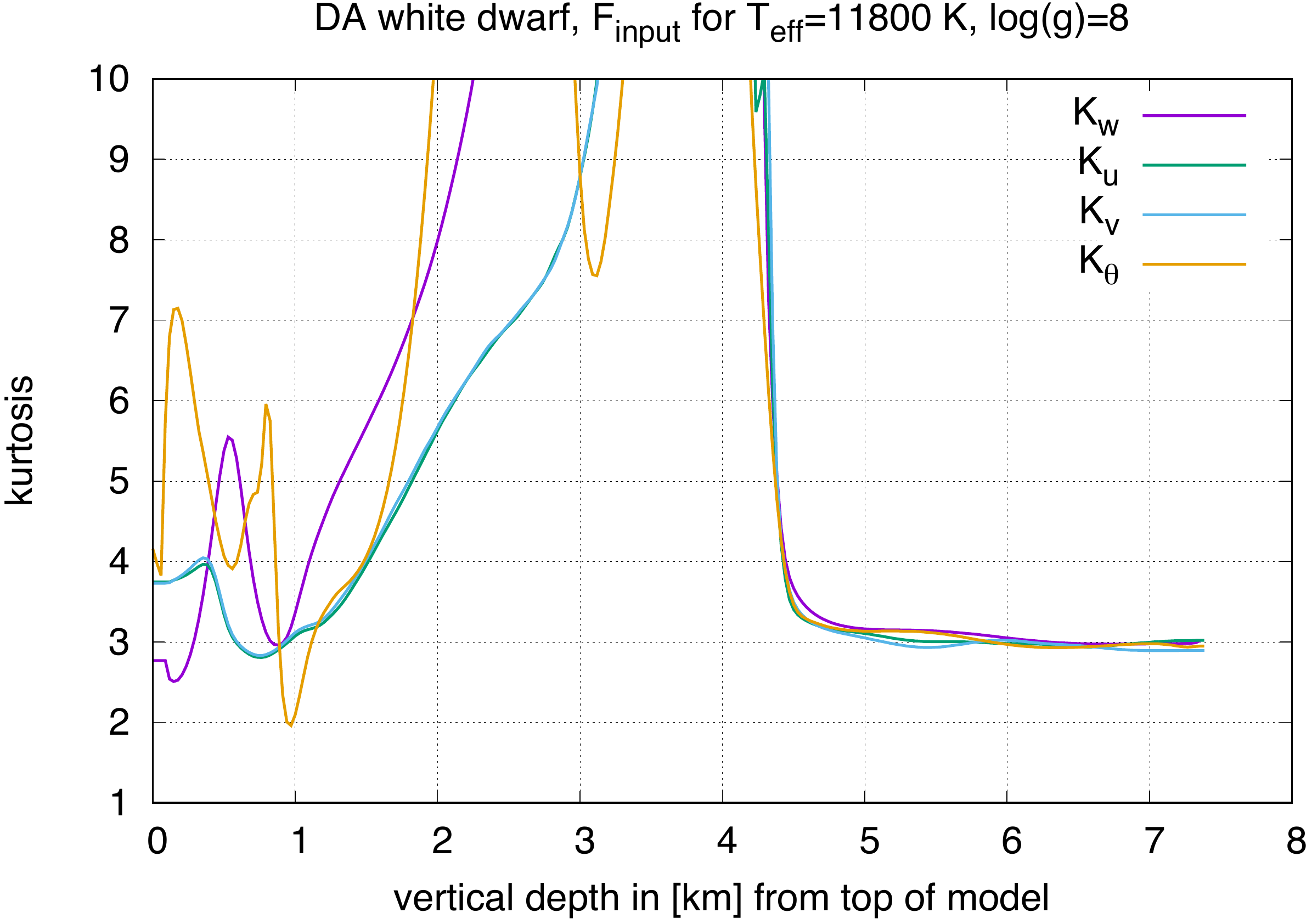}
        \caption{Kurtosis of fluctuations of the temperature field, $K_{\theta}$, and the
                     components of the velocity field, $K_{w}, K_{u}, K_{v}$, around their horizontal mean, averaged over 
                     each layer and in time (over $t_{\rm stat}$), plotted as a function of depth (values truncated above 10).}
    \label{fig:kurtosis_lr}
\end{figure}

In Fig.~\ref{fig:kurtosis} we display the kurtosis of fluctuations of the temperature field, $K_{\theta}$, and 
the vertical and horizontal components of the velocity field, $K_{w}$, $K_{u}$, and $K_{v}$, computed
in analogy to skewness except that now $K_{\theta} = \langle (T-\langle T \rangle_{\rm h})^4\rangle_{\rm h,t} / 
(\langle (T-\langle T \rangle_{\rm h})^2\rangle_{\rm h,t})^{2} =\langle T'^4\rangle_{\rm h,t} / (\langle T'^2\rangle_{\rm h,t} )^{2}$,
and likewise for the components of the velocity field. The kurtosis provides a measure of the strength
and importance of deviations of a fluctuation from a given root mean square average. For a mathematically
meaningful distribution, $K \geqslant 1$. For a Gaussian distribution, $K = 3$ in addition to $S = 0$.
These are necessary though not sufficient conditions for Gaussianity. As one can read from 
Fig.~\ref{fig:kurtosis} the plumes lead to extreme values of $K_w$, especially where a few, isolated 
plumes dominate the distribution (around 4~km). In comparison, for $K_{\theta}$ the flux boundary with
its reversion from cold to hot plumes leads to a local minimum in that region (around 3~km). For
the Sun or other main sequence stars (\citealt{kupka07b,kupka09c}) the network of downflows with 
its embedded granules leads to global minima of $K_{\theta}$ and $K_w$ at the superadiabatic peak and we
find this also for our simulation of a DA white dwarf (neglecting the top of the simulation box)
(see Fig.~\ref{fig:kurtosis_lr}). Likewise, for both $K_{\theta}$ and $K_w$ the
plumes in the overshooting region lead to very large values of K which characterises their large deviation
from the root mean square values of fluctuations of vertical velocity and temperature. In the wave-dominated
region eventually $K \rightarrow 3$ for each of the four fields depicted in Fig.~\ref{fig:kurtosis_lr}. Along with
their values of $S \approx 0$ this demonstrates their different nature in comparison with the plume-dominated 
region further above. We note that the plumes also lead to very large values of $K_u$ and $K_v$ since they 
lead to locally very large velocities (the precise values at  Fig.~\ref{fig:kurtosis} around 4~km are less certain
again due to the limited statistics of a small number of plumes).

\subsection{Properties of the velocity field and mixing}

\subsubsection{Analysis of the velocity field and best fit functions}

With the statistical and topological properties of the velocity and temperature field
in mind we now turn to a more detailed characterization of overshooting and its
associated root mean square velocity fields.

We find an exponential decay of $w_{\rm rms}$ only in a very general sense. 
As we have shown in Sect.~\ref{subsec:velocities}, in the plume--dominated region 
down to the deepest layer where still $F_{\rm conv}/F_{\rm input} < -1\%$, the assumption
of a linear dependence of $w_{\rm rms}$ on $\log T$ allows an accurate fit.
Indeed, its root mean square deviation normalised to the data range is just $\lesssim 1.2\%$ over
a range of $\sim 1.1\, H_{\rm p}$ (Fig.~\ref{fig:logT}). If the fit is made with $T$ as an independent
variable over the same depth region, one even obtains a marginally lower fit error (by 0.07\%). 
The smallness of the difference between the two is due to the limited region in $\log T$ over which
the fit occurs. We expect that from the viewpoint of a physical interpretation, the linear fit in $T$
is physically more relevant. Now if we instead aim at finding the best exponential fit of the velocity 
field which at least partially includes the plume--dominated region and is of identical extent 
($\sim 1.1\, H_{\rm p}$), we have to shift the $\Delta \ln P$ range (see Fig.~\ref{fig:scaling}) to 
the region from 0.35 to 1.45 to obtain a root mean square deviation normalised to data range 
of $\lesssim 1.6\%$ for $\Delta \ln w_{\rm rms}$. This is larger by $1/3$ compared to the
linear fit in $T$. Moreover, as discussed in Sect.~\ref{subsec:velocities}, the fit in 
$\log T$ (or $T$) occurs over {\em exactly the plume--dominated region}, except for
the lowermost part, which is dominated by rare events (single plumes) and even for this
region the fit of $w_{\rm rms}$ versus $\log T$ is at least good for deriving a lower limit for 
the region which can be assumed to be very well mixed by overshooting (down to $\sim 4$~km 
or $\log(1-M_r/M_{\star})\sim -13.9$). In comparison, the clearly poorer exponential fit
begins only around 3.08~km and ends around 4.14~km. Thus, it starts already right inside
the plume-dominated region and includes the physically very different region where plumes
have essentially disappeared. It is hence both less accurate and has a less obvious physical
motivation. 

As mentioned already in the discussion of Fig.~\ref{fig:scaling} in 
Sect.~\ref{subsec:velocities}, one could attempt fitting also the transition regions 
from the plume--dominated to the counter--gradient region and from the plume--dominated 
to the wave--dominated region by exponential fits with different decays. However, these 
hold for even smaller depth ranges and we see little physical motivation for them
beyond providing decent mathematical fits. A ``compromise exponential fit'' for
both transition regions and the plume--dominated region yields a clearly poorer 
result than any of the more localised fits.

We hence suggest that the best approximation for the plume--dominated region
is that of an approximately {\em linear} decay of rms velocities with depth, as it is also 
found from the Reynolds stress model as used by \cite{montgomery04b}. We emphasise 
that this statement is restricted to the case of overshooting zones with strong, hot plumes 
that are subject to high radiative losses. Clearly, this does not include the case where 
convection zones have a marginal $F_{\rm conv}/F_{\rm input}$ already inside
the convective zone itself (i.e., for $T_{\rm eff} \gtrsim 13\,000~{\rm K}$) and to
understand the complex variation of overshooting with parameters such as $T_{\rm eff}$ 
and $\log(g)$, which was studied by \cite{tremblay15b}, requires a grid of models,
as has indeed been used by these authors.  

\subsubsection{The wave--dominated region}   \label{subsubsec:wave-dominated}

How is the region underneath the plume--dominated region different from layers further 
above it? In previous literature on overshooting inside stars, little attention appears 
to have been paid to the horizontal velocity field. As Fig.~\ref{fig:velocities} demonstrates, 
after a transition region of $\sim 0.5$~km below the very well mixed region, we find 
a nearly linear decay of horizontal kinetic energy relative to vertical one (see the 
behaviour of $\Phi$ in Fig.~\ref{fig:logT} and Fig.~\ref{fig:velocities}). This is remarkable, 
since horizontal velocities are expected to {\em increase} relative to vertical ones, if the flow 
approaches a closed, stress-free vertical boundary. $\Phi$ {\em decreases} linearly over 
1.15 pressure scale heights with an error of $\lesssim 1.9\%$. We note that an exponential 
decay at optimised decay rate is slightly better ($\lesssim 1.4\%$). If optimised fits of exponential
decay are individually made for $w_{\rm rms}$, $u_{\rm rms}$, and $v_{\rm rms}$, residual
errors are in the range of $0.5\%$--$0.9\%$. Thus, although the decay of the root mean square velocity fields 
and the quantity $\Phi$ is exponential to very high accuracy, we can use the linear decay of 
$\Phi$ observed in the simulated region as a lower estimate for the extent of mixing. 

We emphasise that the clear identification of this exponential decay for the case of 
overshooting with strong, hot plumes (i.e., at the $T_{\rm eff}$ of our simulation 
target) requires the extra extent of $\sim 1\, H_{\rm p}$ in comparison with earlier work
(\citealt{tremblay11b,tremblay13b,tremblay15b} and even more so \citealt{freytag96b}).
This allows a clear separation of the impact of the lower boundary condition on the
velocity fields, especially on the horizontal components, where this is more 
pronounced (Fig.~\ref{fig:velocities}). The detailed analyses of
exponential decay of velocities underneath the convection zone have focussed
more on hotter DA white dwarfs in earlier work, for which conveniently a sufficiently
deep simulation is more easily achieved, because the equivalence of a plume-dominated 
region cannot exist in their case, as $|F_{\rm conv}|/F_{\rm input} \lesssim 1\%$ for
these stars even inside their convectively unstable zone. Thus, the total number of pressure 
scale heights needed in a simulation to model the wave-dominated region such that at
least the upper part of it is not affected by the lower boundary condition is smaller.

Since in our case, only underneath about 6~km, at still a pressure scale height distance 
from the lower boundary condition, the latter begins to notably influence the flow, we conclude the decay 
of horizontal relative to vertical velocities, which is visible in each of Fig.~\ref{fig:velocities}--\ref{fig:logT},
to be real. Below about 4~km (where $F_{\rm rad} \rightarrow F_{\rm total}$), the flow gradually 
changes into a slow, chiefly vertical, and thus a more wave-like motion. Independently from the 
limitations of even the present simulation, it is clear that a lack of horizontal flow prevents 
mixing. Thus, the mixing in this region is much less efficient than in the layers of notable 
convective energy flux, and the effective diffusivities and mixing time scales are expected 
to differ by a very large amount, since a wave-like motion is much less efficient in entraining
fluid in comparison to an overturning flow.

In the region between 4.45~km and 5.85~km the fluctuations 
of density and temperature decay rapidly ($\lesssim 3\cdot 10^{-3}$ 
of their horizontal mean). Skewness and kurtosis, as depicted in 
Fig.~\ref{fig:skewness}--\ref{fig:kurtosis_lr} of the velocity and temperature 
fluctuations, yield $S = 0 \pm 0.06$ and $K = 3 \pm 0.2$ for all layers below 
4.85~km (at 3.5~km we find $K \gg 3$, $|S| > 1$). This excludes a flow with 
a granulation pattern or thin plumes embedded in gently moving upflows,
as is proven by Fig.~\ref{fig:T_slices}. 
The components and magnitude of vorticity of $\mathbf{u}$ drop faster than 
$w_{\rm rms}$, $u_{\rm rms}$, and $v_{\rm rms}$ in that region. No indications 
for ``extreme events" have appeared during $t_{\rm stat}$. 

\begin{figure}
	%\includegraphics[width=\columnwidth]{Mode_P_0p263s}
	\includegraphics[width=\columnwidth]{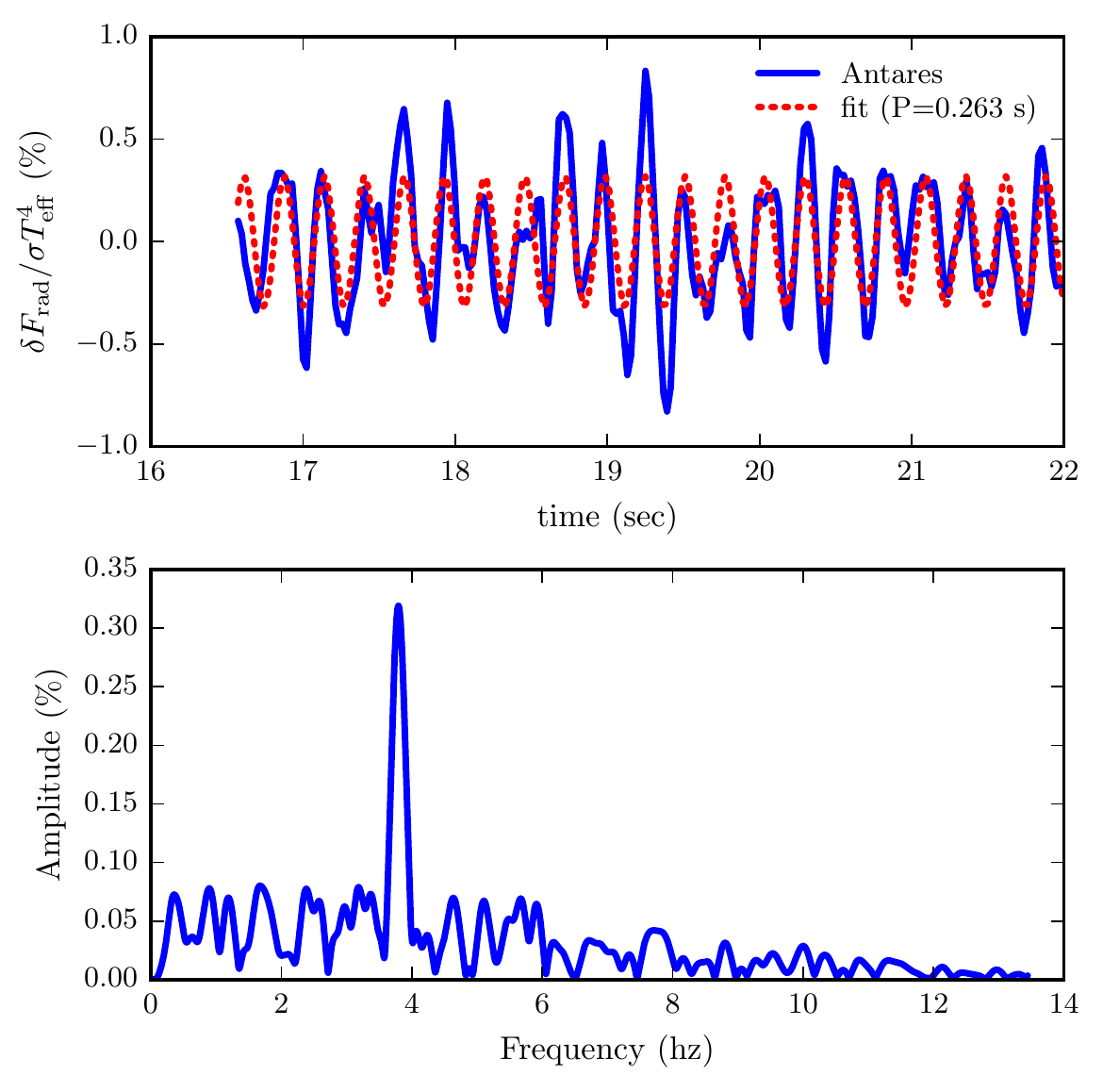}
        \caption{Pressure oscillation mode in our numerical
          simulation, whose period is close to the sound
          crossing time. In the upper panel we show the fractional
          variation of the radiative flux emerging at the top of the
          simulation domain after division by a polynomial fit to
          remove the trend due to thermal relaxation (blue, solid
          curve); the dotted (red) curve shows the sinusoidal fit  to
          these data. The lower panel gives the Fourier transform of
          the normalized flux variations, which indicates a mode with
          a period of 0.263~s and an
          amplitude of $\sim 0.3$\%.
          }
    \label{fig:pmode}
\end{figure}

We also observe a stable oscillation, a vertical, global pressure 
mode without interior node, which is easily visible for the horizontally averaged, 
vertical mean velocity. It has a frequency of $\nu_{\rm osc} = 3.7959\,{\rm Hz}$
and thus a period of $P_{\rm osc} = 0.26344\,{\rm s}$. 
Its amplitude is $\sim 30~{\rm m\, s}^{-1}$ at $\sim 5.85$~km where 
$w_{\rm rms} \sim 105~{\rm m\, s}^{-1}$. It can also be identified in 
the emerging radiative flux even though in this case it is subject to
perturbations by local events (shock fronts, etc.) and drift due to thermal relaxation
(discussed in the context of Fig.~\ref{fig:relax-fluxes}). After removing the latter
from $F_{\rm rad}(x=0\, {\rm km},t)$ by a polynomial fit, it is easy to extract the dominant 
mode frequency, as is demonstrated in Fig.~\ref{fig:pmode}. Once excited, this oscillation cannot 
be removed from the simulation by artificial damping, if that were attempted (we have done a number 
of experiments with 2D simulations to corroborate that), so evidently the convection 
zone in the simulation is able to feed the mode energetically at a sufficiently high 
rate against damping mechanisms (due to radiative losses, viscosity, etc.)
to support this very stable amplitude. The mode is also present throughout
the convection zone and even has its maximum amplitude right there where
it provides just a small fraction of the kinetic energy of the flow.

The oscillation with a frequency of $\nu_{\rm osc}$ can also be identified when 
tracing hot or cold structures in Fig.~\ref{fig:T_slices}h--i, in $w_{\rm rms}$, 
or in $F_{\rm rad}$ (see Fig.~\ref{fig:relax-fluxes}). Thus, the flow taking
place in the region below 4~km is a combination of global waves and local, transient 
features. Its properties with respect to skewness, kurtosis, vorticity,
the clear presence of a global vertical mode with a significant contribution 
($\sim 10\%$) to the kinetic energy in that region, and the decrease of 
$\Phi$ (and thus the faster decrease of horizontal in comparison
with vertical velocities) justifies its naming as {\em wave--dominated region}. 
But these properties are at variance with efficient mixing and we thus expect 
the mixing processes in the two regions, the overshooting zone proper and
the wave--dominated region, to differ physically and in their efficiency.

Now one might use the velocity field found in the numerical simulation,
as has been done before (\citealt{freytag96b}, \citealt{tremblay15b}, e.g.),
and derive a depth below which diffusion velocities dominate and in this sense 
define an {\em extent of the mixed region}. We have to point out here
a major caveat of this procedure: just as simulations of solar surface
convection, which have open lower vertical boundaries, the solid, slip boundaries
used for simulations with a radiative region at the bottom also 
{\em reflect vertical waves}. Thus, while fluid can leave or enter
the domain only in the former case, {\em both} types of boundary conditions conserve 
momentum inside the box {\em and} create a reflecting layer for vertical waves at the 
bottom of the simulation box. For both cases waves in a real star should have smaller 
maximum amplitudes due to the simple fact that the {\em mode mass} contained 
in the simulation box is much smaller. Amplitudes of p-mode oscillations hence have 
to be scaled by mode mass \citep{stein2001b} to be compared to observations
(or models of the entire star). Although the theoretical explanation of
waves excited in the overshooting zone proposed by \cite{freytag96b}
is based on ${\rm g}^{-}$-modes rather than p-modes, we expect the former
also to be altered (towards exhibiting a much lower equilibrium amplitude)
due to the fact that in a real star such waves connect to a much larger mass.
Hence, we expect that the velocities obtained for the wave--dominated
region are systematically overestimated compared to a model of a
full star. A determination of the mixed region by means of $w_{\rm rms}$ then 
leads to an even more pronounced overestimation. Consequently, 
as an upper limit of the mixed region one should rather use $u_{\rm rms}$
and $v_{\rm rms}$ or an exponential fit of $\Phi$. 

Having the inevitable overestimations by this procedure in mind
and without a suitable method of scaling of the velocity fields yet at hands,
we may also consider the linear fit of $\Phi$ which at least 
provides a lower estimate for mixing due to the processes in
the wave--dominated region, if the velocities of the numerical
simulation in that part are taken at face value. Indeed, if we extrapolate the 
linear fit of $\Phi$, it would reach a value of 1 at $\sim 7.67$~km, some 
5.67~km below the lower boundary of where $\nabla > \nabla_{\rm ad}$ still holds. 
This hopefully provides a safe, lower limit for the extent of the well mixed region; it 
corresponds to a mass of $\log(1-M_r/M_{\star})\sim -12.7$ in our stellar model.

\subsubsection{Mixing and accretion}

For the case of steady-state accretion of metals, the observed surface
abundance results from a competition between the accretion rate onto
the surface and the settling rate of the metals at the base of the
mixed region. Using the published values of \citet[][Tables 1 and
4]{koester09}, we can estimate the effect that mixing beneath the
formally convective region has on the settling rates for trace amounts
of metals in a hydrogen atmosphere white dwarf. If we assume that
mixing only occurs in the region defined by the Schwarzschild
criterion, the base of the mixed region would be at
$\log(1-M_r/M_{\star})\sim -15.2$; interpolating in Tables 1 and 4 of
Koester yields a settling time for carbon of
$\tau_{-15.2} \sim 0.14\,$yr. On the other hand, if we assume the base
of the mixed region is at $\log(1-M_r/M_{\star})\sim -13.9$,
corresponding to the depth of penetration of the plumes, then we find
the settling time of carbon is $\tau_{-13.9} \sim 1.4\,$yr.  Finally,
if we take the base of the mixed region to be
$\log(1-M_r/M_{\star})\sim -12.7$ (where $\Phi = 1$ is extrapolated
from the ANTARES simulation), then we obtain a settling time for carbon
of $\tau_{-12.7} \sim 12\,$yr. The ratio of this settling time to that
assuming mixing only in the Schwarzschild unstable region is
$\sim 87$.  For the elements Na, Mg, Si, Ca, and Fe, we find similar
ratios for the enhancement of their settling times, in the range of
50--97.  Thus, including the mixing in the overshooting region has a
very large effect on the computed settling times of metals in WD
envelopes.

In a similar vein, we would like to examine the effect that a larger
mixed region has on the inferred accretion rates, assuming
steady-state accretion. From Eq.~6 of \citet{koester09},
$X_{\rm CZ} = \tau_{\rm CZ} \dot{M}_X (M_{\rm CZ})^{-1}$, where
$X_{\rm CZ}$ is the mass fraction of element $X$ in the convection
zone, $\tau_{\rm CZ}$ is the settling time at the base of the mixed
region, $\dot{M}_X$ is the mass accretion rate of element $X$, and
$M_{\rm CZ}$ is the mass of the mixed region. If we assume that
$X_{\rm CZ}$ is fixed by the observations, then
$\dot{M}_X \propto M_{\rm CZ}/\tau_{\rm CZ}$.  Using
$\log(1-M_r/M_{\star})\sim -15.2$ and $-13.9$ for the extent of the
mixed region with and without overshooting, respectively, we find that
including the overshooting region enhances the inferred accretion
rates by factors of 1.6--2.5 for the set of metals previously
considered. If instead we take the depth of the overshooting region to
be at $\log(1-M_r/M_{\star})\sim -12.7$, then we find that the
inferred accretion rates are enhanced by factors 3.2--6.3 for the same
set of metals with respect to the no-overshooting case.

We point out here that the total mixed mass obtained from the linear
fit of $\Phi$ with the requirement $\Phi \rightarrow 1$ is roughly 300
times larger than that contained in and above the
Schwarzschild-unstable region.  This number is similar to suggestions
by \cite{freytag96b}, \cite{koester09}, and within the (much larger)
range suggested by \cite{tremblay15b}.

\section{Discussion and outlook}    \label{sec:discussion}

We have highlighted here that estimates for the extent of the mixed
region underneath the surface convection zone of DA white dwarfs
at intermediate effective temperatures ($T_{\rm eff} \approx 11\,800~{\rm K}$)
as obtained from (3D) hydrodynamical simulations have to be revisited
from a new perspective. This has become possible thanks to
the larger (3D) simulation domain and the simulation being performed
over a sufficient amount in time. The larger horizontal extent reduces
artifacts by the periodic boundary conditions and allows a better 
sampling of statistical data due to a larger number of realizations.
The larger vertical extent allows for the first time a clear identification
of exponential decay of the velocity field, as a function of depth, {\em underneath} 
a region dominated by plumes. In the latter, exponential fits lack both accuracy, 
since the decay rate itself would have to be a {\em function} of depth, and also a
physical basis, since the upper parts of the overshooting zone are
completely dominated by plumes and their dynamics instead of
waves (even though the latter are of course also present in that
part of the simulation domain). In the plume--dominated region
(once $F_{\rm conv} < 0$) we find a linear decay with depth to provide 
a more accurate model and in this sense the simulations {\em agree} 
qualitatively and, roughly, quantitatively with solutions of Reynolds stress 
models for somewhat hotter objects (cf.\ \citealt{montgomery04b}). The 
wave--dominated region, for which we confirm exponential decay
of the vertical root mean square velocity, is {\em not} modelled
by the Reynolds stress approach (as the time dependent mean
velocity was assumed to be zero, so it cannot be included in
the model used by \citealt{montgomery04b}). Viewing our results
at a more coarse level we also confirm some basic findings of earlier studies 
(especially from \citealt{freytag96b} and \citealt{tremblay11b,tremblay13b,tremblay15b})
performed with different simulation codes and with smaller domain sizes.

Contrary to earlier work we stress here that the horizontal
velocities, which decay more rapidly than the vertical ones 
in the wave--dominated region, provide an indication for a less
efficient mixing. We also point out that the amplitudes due to waves
obtained from this class of simulations should be expected to be
systematically too large. Thus, while the convective mixing due to
overshooting up to and including the plume--dominated region
is on safe ground, its extension into the wave--dominated
region is much less certain and certainly subject to gross
overestimation. We thus provide a linear extrapolation
based on horizontal velocities which might be used as a more
conservative estimate of the extent of the convectively mixed region.

We emphasise here that our results have been obtained
for the case of a DA white dwarf with a large amount of
convective flux inside the unstable zone which features strong,
long-lived plumes penetrating into the stably-stratified
region. Further studies should clarify the dependencies of these results
on $T_{\rm eff}$ and $\log(g)$, and, with respect to the wave-dominated
region, on simulation depth and time.

Finally, we show that the extent of the mixed region can have 
a large effect on the computed settling times and accretion rates of 
metals in WDs also when using our more conservative estimates
of the extent of mixing due to overshooting. While we think there is merit in
the prescription we have adopted for defining the extent of the mixing
region, it is probably a lower limit compared to results which
assume mixing velocities to decrease exponentially with depth.
More precise estimates of this kind clearly require further work,
not only from the viewpoint of simulations, but also with respect
to some theoretical aspects such as mixing efficiency of the
encountered types of flows and characterizations of
the effect of limited simulation domain sizes.

\section*{Acknowledgements}

F.~Kupka gratefully acknowledges support through Austrian Science Fund
(FWF) projects P25229 and P29172. Parallel simulations have been
performed at the Northern German Network for High-Performance
Computing (project number bbi00008) and the Heraklit cluster at the
BTU Cottbus-Senftenberg.  M.~H.~Montgomery gratefully acknowledges
support from the United States Department of Energy under grant
DE-SC0010623 and the National Science Foundation under grant
AST-1312983.

%%%%%%%%%%%%%%%%%%%%%%%%%%%%%%%%%%%%%%%%%%%%%%%%%%


\bibliographystyle{mnras}
\bibliography{letter_wd_11800} % if your bibtex file is called example.bib

%%%%%%%%%%%%%%%%%%%%%%%%%%%%%%%%%%%%%%%%%%%%%%%%%%

%%%%%%%%%%%%%%%%% APPENDICES %%%%%%%%%%%%%%%%%%%%%

%\appendix


%%%%%%%%%%%%%%%%%%%%%%%%%%%%%%%%%%%%%%%%%%%%%%%%%%

% Don't change these lines
\bsp	% typesetting comment
\label{lastpage}
\end{document}